\documentclass{emulateapj}

\usepackage{graphicx}

\newcommand{\kms}{\rm km\ s^{-1}}

\newcommand{\vecv}{{\rm\bf{v}}}
\newcommand{\vecB}{{\rm\bf{B}}}
\newcommand{\vecnabla}{{\bf{\nabla}}}
\newcommand{\HI}{H\,{\small I}}
\newcommand{\HII}{H\,{\small II}}

\submitted{Accepted for publication in The Astrophysical Journal}

\shorttitle{3D Simulations of Superbubbles}
\shortauthors{Stil et~al.}

\begin{document} 
\title{Three-Dimensional Simulations of Magnetized Superbubbles:\\
New Insights into the Importance of MHD Effects on Observed Quantities\\}
\author{{\sc Jeroen Stil, Nicole Wityk, Rachid Ouyed, and A. R. Taylor}}
\affil{Department of Physics and Astronomy, University of Calgary, 2500 University Drive NW, Calgary, Alberta, T2N 1N4 Canada}

\begin{abstract}
We present three-dimensional magnetohydrodynamic (MHD) simulations of
superbubbles, to study the importance of MHD effects in the
interpretation of images from recent surveys of the Galactic
plane. These simulations focus mainly on atmospheres defined by an
exponential density distribution and the \citet{1990ARA&A..28..215D}
density distribution. In each case, the magnetic field is parallel to
the Galactic plane and we investigate cases with either infinite scale
height (constant magnetic field) or a constant ratio of gas pressure
to magnetic pressure.  The three-dimensional structure of superbubbles
in these simulations is discussed with emphasis on the axial ratio of
the cavity as a function of magnetic field strength and the age of the
bubble.  We investigate systematic errors in the age of the bubble and
scale height of the surrounding medium that may be introduced by
modeling the data with purely hydrodynamic models. Age estimates
derived with symmetric hydrodynamic models fitted to an asymmetric
magnetized superbubble can differ by up to a factor of four, depending
on the direction of the line of sight. The scale height of the
surrounding medium based on the Kompaneets model may be up to 50\%
lower than the actual scale height.  We also present the first ever
predictions of Faraday rotation by a magnetized superbubble based on
three-dimensional MHD simulations.  We emphasize the importance of MHD
effects in the interpretation of observations of superbubbles.
\end{abstract}

\section{INTRODUCTION}
\label{intro}

The combined stellar wind and supernova ejecta of groups of O and B
stars blow large bubbles in the interstellar medium. The largest of
these bubbles, with size scales of 100 pc to 1 kpc are commonly
referred to as superbubbles. The basic structure of a superbubble
consists of a hot low-density interior, the cavity, surrounded by a
cool shell of swept-up interstellar medium. The continuous formation
and dissipation of superbubbles is an important factor in the energy
balance of the interstellar medium, and determines the locations of
different phases of the interstellar medium on large scales
\citep{mckee1977}. Compression of the interstellar medium in the shell
may increase cooling and trigger the formation of a new generation of
stars.  Also, the ability of large superbubbles to break out of the
disk of a galaxy and initiate an outflow of chemically enriched plasma
and ionizing radiation from the disk into the halo has a profound
influence on the evolution of galaxies.

Several examples of well-defined superbubbles have been identified in
the Galaxy (e.g. Heiles 1984; Maciejewski et~al. 1996; Normandeau et
al. 1996; Heiles 1998; Ehlerov\'a \& Palous 1999; Callaway
et~al. 2000; Reynolds et~al. 2001; McClure-Griffiths et al. 2002, Pidopryhora
et~al. 2007). These have been studied in detail, thanks to their
relative proximity. Observations with parsec-scale resolution of
neutral and ionized gas reveal important details about the interaction
between the hot ejecta and the interstellar medium.  New
high-resolution surveys of Galactic atomic hydrogen (\HI) emission
\citep{taylor2003,mcclure2005,stil2006} have provided unprecedented
images with morphological and kinematic information of Galactic
superbubbles \citep{1996Natur.380..687N,mcclure2003}.  Physically
interesting parameters are usually derived from the observations by
means of analytic models that assume spherical symmetry
\citep{1975ApJ...200L.107C,1977ApJ...218..377W} or axial symmetry
\citep{Kompaneets1960,1999ApJ...516..843B}.

In this paper we investigate the importance of MHD effects on physical
quantities derived from observed superbubbles. The effect of the
Galactic magnetic field is difficult to model because it introduces
anisotropy in the medium, which requires three-dimensional numerical
simulations. The first three-dimensional magnetohydrodynamic
simulations of superbubbles were presented by
\citet{1998MNRAS.298..797T}, who discussed the importance of the
magnetic field in the break-out of superbubbles from the Galactic
disk. \citet{1998MNRAS.298..797T} also described significant
departures from spherical and axial symmetry in the shape of a
magnetized bubble resulting from the interaction of the expanding
superbubble with the Galactic magnetic field. The shape and the size
of a superbubble depend on the strength and the geometry of the
Galactic magnetic field as much as they depend on the density
distribution of the ambient interstellar medium. The expanding
superbubble in turn redefines the geometry of the interstellar medium
and the Galactic magnetic field in a volume several hundred parsecs
across.

\citet{korpi1999} and \citet{deavillez2005} performed thee-dimensional
MHD simulations that include the evolution of a superbubble in a
supernova-driven turbulent multi-phase interstellar medium. The super
bubbles in these simulations show significant departures form symmetry
because of inhomogeneities in the medium in which the super bubble
expands. \citet{korpi1999} and \citet{deavillez2005} found that a
super bubble can break out into the halo in such a medium. Although
these simulations provide valuable insight in the dynamics of
superbubbles in a multi-phase interstellar medium, it is in general
difficult to relate these simulations to specific observed super
bubbles (see however Fuchs et~al. 2006 for a simulation of the Local
Bubble).

Astrophysical parameters derived from observations, such as the age
and the energy of a superbubble, or the density distribution of the
ambient medium have relied on symmetric analytic models that do not
include a magnetic field.  The departure from axial symmetry imposed
by the magnetic field introduces systematic errors that have not been
considered before. In this paper we present three-dimensional MHD
simulations of superbubbles, and we explore the errors introduced by
commonly used methods to determine basic parameters from
observations. In particular, we study the axial ratio of the
wind-blown cavity for different times and magnetic field
configurations.  The smaller simulation volume and time span used in
our simulations compared to \citet{1998MNRAS.298..797T} are more
tailored to the latitude coverage of the Galactic plane surveys. We
also calculate the first images of Faraday rotation by a magnetized
superbubble derived from our simulations, and emphasize the importance
of such simulations to make meaningful predictions in this area.

We describe frequently used analytic models \S 2 and detail our
numerical setup and methods in \S 3.  Numerical results and the
analysis of magnetic effects on derived parameters are presented \S 4.
The specific case of the W4 supperbubble is discussed in \S 5. Faraday
rotation by magnetized superbubbles is discussed in \S 6, and
conclusions are presented in \S 7.

\section{ANALYTIC HYDRODYNAMIC MODELS OF BUBBLES}
\label{analytic}
The exact shape and size of the bubble depends directly on the
environment and the strength of the source. For weaker sources which
produce bubbles whose extent are much smaller than the scale height of
the gas in the galaxy, a solution with an approximately constant
density profile may be considered. \citet{1975ApJ...200L.107C} present
a solution for a spherical wind-blown bubble expanding into a
non-magnetized medium, in which the radius of the bubble varies
according to
%\begin{equation}
\begin{eqnarray}
R_{b}(t) = 0.76 \left(\frac{L_{s}t^{3}}{\rho}\right)^{1/5} = \nonumber \\
55.2\left(\frac{L_{s}}{3\times 10^{37}\ \rm erg\ s^{-1} }\right)^{1/5} \times \nonumber \\
\times \left(\frac{\rho}{1.67\times 10^{-24}\ \rm g\ cm^{-3}}\right)^{-1/5}
t_6^{3/5}\ \rm pc.
\label{rcastor}
\end{eqnarray}
%\end{equation}
where $R_{b}$ is the outer shock of the shell, $L_{s}$ is the
mechanical luminosity of the source, $\rho$ is the ambient density of
the undisturbed interstellar medium, and $t_{6}$ is the age of the
bubble in Myr. The radius of the contact discontinuity was found 
to be $0.86 R_b$ by \citet{1977ApJ...218..377W}.
   
For larger bubbles, associated with powerful sources, the situation is
more complicated.  The gravitational potential of the Galactic disk
produces a density gradient perpendicular to the plane of the disk. At
early stages, the \citet{1975ApJ...200L.107C} solution can still be
applied, but as the radius of the bubble exceeds the scale height of
the surrounding medium, the density gradient of the surrounding medium
affects the shape of the bubble.

\citet{Kompaneets1960} (hereafter K60) proposed a solution for the
evolution of a point explosion in a non-magnetized exponential
atmosphere that can be applied to this situation. This solution is
based on the assumptions that the thermal energy is a constant
fraction of the energy deposited in the initial blast and that the
energy distribution is uniform throughout the volume of the bubble
except close to the shock front where the energy density can be two to
three times the mean value (see also Bisnovatyi-Kogan~\&~Silich
1995). \citet{1999ApJ...516..843B} extended the Kompaneets solution to
describe continuous energy injection, which we use in this paper. The
source is located at $z=0$ in an exponential atmosphere with scale
height $H$ of the form
\begin{equation}
\rho(z) = \rho_{o}e^{-z/H}.
\end{equation}

The radius of the bubble in cylindrical coordinates takes the
following form
\begin{equation}
R = 2H\arccos\left[\frac{1}{2}e^{z/2H}\left(1 - \left(\tilde{y}/2\right)^{2} + e^{-z/H}\right)\right]
\label{rbasu}
\end{equation}
where 
\begin{equation}
\tilde{y} =\frac{1}{H} \int_o^t\sqrt{\frac{\gamma^{2} - 1 }{2}\frac{E_{th}}{\rho_{0}\Omega}}dt'
\label{ybasu}
\end{equation}
is a dimensionless transformed time variable, $\gamma$ is the
adiabatic index, $E_{th}$ is the thermal energy of the bubble, and
$\Omega$ is the volume of the bubble.

The independent variables are the mass density at $z=0$ ($\rho_{0}$),
the scale height of the atmosphere ($H$), and the mechanical
luminosity of the source ($L_{s}$), which determines $E_{th}$. The
unit of time in this model is then defined by
\begin{equation}
t_{0,k}=\left(\frac{\rho_0 H^{5}}{L_s}\right)^{\frac{1}{3}}. 
\label{ytilde}
\end{equation}

For $R=0$, Equation (\ref{rbasu}) reduces to two equations for the
upper and lower boundary of the blast wave, $z_1$ and $z_2$
respectively \citep{1999ApJ...516..843B}
\begin{equation}
\label{zupperlower}
z_{1,2} = -2H \ln\left( 1 \mp \frac{\tilde{y}}{2}\right) 
\end{equation}

As $\tilde{y}$ approaches 2, the top of the bubble in the Kompaneets model
expands to infinite height in a finite amount of
time. Physically, this means the shock acceleration in the
z-direction becomes infinite because of the strong density gradient
\citep{Bisnovatyi1995}. The bottom of the bubble ($z_{2}$, Equation
\ref{zupperlower}) does not penetrate downward more than $2 H \ln 2
\approx 1.4 H$, its location at the time of blow-out. Since
the top of the Kompaneets model reaches an infinite height in a finite
amount of time, it cannot be a valid solution at later
times. However, it can provide an adequate solution at early times, if
the initial conditions are consistent with the assumptions of the
Kompaneets model, i.e. negligible pressure of the ambient medium and
negligible inertia of the swept up medium.

With the addition of a magnetic field, no three-dimensional analytic
solutions exist. In order to capture the true evolution of the these
bubbles numerical simulations need to be performed.

\section{SIMULATION SETUP}

\subsection{Goals and Limitations}
 
We aim to include the physics of the Galactic magnetic field in the
interpretation of observed superbubbles. The derivation of physical
quantities from the data, such as the scale height of the surrounding
medium or the age of a superbubble is best served by a model that
takes into account the magnetic field, but not the complications of a
preprocessed interstellar medium.  We do not include the effect of
Galactic differential rotation or a Coriolis force on the superbubble
in our simulations.  \citet{1998MNRAS.298..797T} found that the
characteristic time scale for shear from differential Galactic
rotation is $\sim$320 Myr while the time scale for shear due to
rotation of the bubble by the Coriolis force is $\sim$50 Myr. These
processes operate over significantly longer times scales than those
considered here (up to 20 Myr).

The evolution of the bubble will also be affected by heating and
cooling processes. In this paper we discuss adiabatic simulations as
an approximation of the situation where heating balances cooling
throughout the life of the superbubble, while we use simulations with
cooling to explore how the evolution changes if cooling dominates over
heating.  Observations of ionized gas associated with superbubble
shells indicate that a significant fraction of the mass in the shell
may be photo-ionized by the central star cluster and the surrounding
interstellar radiation field. The ionized mass in the shell of the
Orion-Eridanus superbubble is $7 \times 10^4 d_{400}^2.5\ \rm M_\sun$
\citep{reynolds1979}. The neutral mass of this shell was determined to
be $5.2 \times 10^5 d_{400}^2\ \rm M_\sun$ by \citet{heiles1976}, and
$2.5 \times 10^5 d_{400}^2\ \rm M_\sun$ by \citet{brown1995}. These
values indicate that 13 to 28\% of the mass of the Orion-Eridanus
shell is ionized by the stars inside the
bubble. \citet{pidopryhora2007} found equal amounts of neutral and
ionized gas in the Ophiuchus superbubble, that is possibly ionized
from the outside as well as from the inside.  The shell of the W4
superbubble \citep{1996Natur.380..687N} is clearly visible by its
thermal radio continuum emission and $H\alpha$ emission
\citep{dennison1997}, but not in \HI. Photo-ionization of a
substantial fraction of the mass of the shell indicates that heating
by photoionization is a non-negligible term in the energy budget of
the gas in the shell.  Simulations that include cooling without
photo-ionization therefore neglect a significant heating term in the
energy budget of the shell. For computational reasons we cannot solve
radiative transport in the three-dimensional MHD simulations. The
absence of cooling in our current simulations corresponds with the
approximation of equilibrium between heating and cooling in the shell.
In Section~\ref{cooling-sec} we discuss the effect of cooling on the
axial ratios of the cavity.

\begin{center}
\begin{deluxetable}{cccc}
\tablecolumns{4}
\tablewidth{0pt}
\tablecaption{Dimensionless Parameter Conversion \label{dimlesstable}}
\tablehead{
\colhead{Parameter}&\colhead{Variable}& \colhead{}& \colhead{Conversion}}
\startdata
Density&$\rho$& & $\tilde{\rho}\rho_{o}$\\
Location&$x$& & $\tilde{x}$H\\
Velocity&$\vecv$& & $\tilde{\vecv}c_{s}$\\
Time&$t$& & $\tilde{t}H/c_{s}$\\
Luminosity & $L$ & & $\tilde{L} \rho_{o}H^{2}c_{s}^{3}$\\
Pressure & $p$ & & $\tilde{p} \rho_{o}c_{s}^{2}$\\
Internal Energy & $e$ & &$\tilde{e} \rho_{o}c_{s}^{2}$\\
Magnetic Field& $\vecB$ & &$\tilde{\vecB}\beta^{-1/2}\rho_{0}^{1/2}c_{s}$ \\
\enddata
\end{deluxetable}
\end{center}
\begin{center}

\subsection{Basic Equations}

\begin{deluxetable}{cccc}
\tablecolumns{3}
\tablewidth{0pt}
\tablecaption{Input Parameters \label{freepar}}
\tablehead{
\colhead{Atmosphere} & \colhead{}& \colhead{}& \colhead{Source}}
\startdata
$\rho_{0}$                                &&& $L_{s}$\\
$\beta_{0}$                               &&& $v_{s}$\\
$\textbf{B}$:(B$_{1}$,B$_{2}$,B$_{3}$)\tablenotemark{a}    &&& $R_{s}$\\
Density profile\tablenotemark{b}            &&& $t_{on}$\\
                                          &&& $t_{off}$\\
                                          &&& Location:($x_{1}$,$x_{2}$,$x_{3}$)\\
\enddata
\tablenotetext{a}{In the simulations presented here the initial magnetic field is oriented along the $x_{1}$ axis ($B_{2}=B_{3}=0$), parallel to the Galactic plane.}
\tablenotetext{b}{Exp $\equiv$ Exponential according to Equation~\ref{rho_exp}; DL $\equiv$ Dickey \& Lockman (1990) according to Equation~\ref{DLeqn}, both with density gradient along the $x_{2}$ axis}
\end{deluxetable}
\end{center}

\label{models}
Numerical simulations of bubbles have been performed by many
researchers (see \citet{2004ApJ...601..621R}, \citet{komljenovic1999},
\citet{1998MNRAS.298..797T}, \citet{1993ApJ...409..663M},
\citet{MacLow1989} and \citet{1986PASJ...38..697T}, among
others). \citet{1998MNRAS.298..797T} was the first to perform
three-dimensional MHD simulations with radiative cooling assuming
symmetry with respect to the $x$=0, $y$=0 and $z$=0 planes. In the
simulations we present here, we have a complete three-dimensional
bubble evolving in an unperturbed, three-dimensional environment.

Our simulations solve the following equations:
\begin{eqnarray}
\frac{\partial\rho}{\partial t} + \vecnabla \cdot (\rho\ \vecv) &=& 0\\
\frac{\partial\vecB}{\partial t} - \vecnabla \times (\vecv \times \vecB) &=& 0\\
\rho\left[\frac{\partial\vecv}{\partial t} + (\vecv \cdot \vecnabla)\vecv\right] - \vecnabla p - \rho\vecnabla\Phi + \nonumber\\
+\frac{(\vecnabla \times \vecB) \times \vecB}{4\pi} &=& 0\\
\label{a}
\rho\left[\frac{\partial e}{\partial t} + (\vecv \cdot \vecnabla)e\right] + p\ (\vecnabla \cdot \vecv) &=& 0\\
\vecnabla \cdot \vecB &=& 0
\end{eqnarray}
where $\rho$ is the density, $p$ is the gas pressure, $\vecB$ is the
magnetic field, $\vecv$ is the velocity, $\Phi$ is the gravitational
potential, and $e$ is the internal energy of the gas.

For our simulations we work in units of density ($\rho_{0}$), scale
height ($H$) and sound speed ($c_{s}$). The conversion to dimensionless
variables takes the following form
\begin{equation}
\rho = \tilde{\rho}\rho_{o}.
\label{dimlessrho}
\end{equation}
The other parameters are converted in a similar manner (see
Table~\ref{dimlesstable}).  The unit of time in our simulations is
\begin{equation}
t_{0} = \frac{H}{c_{s}}
\label{dimlesstime}
\end{equation}
which is different from the unit of time in the Kompaneets model
$t_{0,k}$ in Equation~(\ref{ytilde}), which assumes $c_{s}=0$.

The MHD equations in dimensionless variables are solved numerically by
the ZEUS-MP code \citep{2000RMxAC...9...66N}. The ZEUS-MP code used in
this work is an adapted and modified version of the original 1.0b
version of Norman and Li (see Norman 2000). One of the authors
involved in this paper (R. Ouyed) has also spent a large amount of time to
modify the code. Most of these modifications were similar to those
described in \citet{Vernaleo2006}.  A detailed description of the
original ZEUS code can be found in Stone \& Norman (1992a\&b) which
also includes basic tests of the code. Vulnerabilities in the ZEUS
family noted by \citet{Falle2002}, in particular the issue of
rarefaction waves and shock errors, have recently been discussed by
Hayes et al. (2006; and references therein). While these papers
acknowledge that the code possess limitations as do all numerical
methods, results from ZEUS-MP were found to compare quite favorably
with other numerical techniques. We come to the same conclusion in
\S~\ref{DLCBandE} where we compare our results to that of
\citet{1998MNRAS.298..797T} despite differences in numerical methods
used.

ZEUS solves the MHD equations using the operator split method.  The
equations are solved in two substeps, called a source step and a
transport step (see Stone \& Norman 1992a\&b). Three methods for the
advection of mass, momentum and internal energy in the transport step
can be implemented: the first order accurate donor cell method
\citep{Godunov1959}, the second order van Leer method
\citep{vanLeer1977}, and the third order piecewise parabolic advection
(PPA) method \citep{ColellaWoodward1984}. After some numerical
experiments we decided to use the van Leer method because it offers
the best ratio of precision to computational costs. The basic
equations of the code are written in a covariant form which allows for
the use of the code in an arbitrary orthogonal coordinate system
(Cartesian, cylindrical and spherical coordinates are predefined). The
algorithm used to guarantee that $\nabla \cdot B=0$ is the ``HSMOCCT''
method which combines the constrained transport (CT) module of
\citet{EvansHawley1988}, and improvements of the method of
characteristics (MOC) introduced by \citet{hawley_stone1995}. In this
scheme, if the initial B has zero divergence in the discretisation on
the staggered mesh, then every time step will maintain the initial
value of the divergence to the accuracy of machine round-off error.

The finite difference method is based on the discretization of each
dependent variable over the spatial computational domain. Then finite
difference approximations to the differential equations are solved on
this discrete mesh. The ZEUS code uses a staggered mesh built up of
two mutually shifted grids. The a-grid specifies positions of the zone
boundaries while the b-grid specifies the zone centers. Discrete
values of all dependent variables are stored for each zone. Scalars
are stored at the zone centers while components of vectors are stored
at the appropriate zone interfaces (see Figure~1 in Stone \& Norman
1992a).  Boundary conditions are implemented as two layers of ghost
zones at each boundary of the computational domain (two layers are
required for higher order interpolation if the PPA method is
used). Values of the dependent variables in the ghost zones are given
by simple, explicit equations that connect these values to the values
in the adjacent active zones. For our simulations we use the so-called
outflow boundary conditions where the values of all variables in the
ghost zones are set equal to the values in the corresponding active
zones.

The simulations are
performed in a three-dimensional cartesian box with right handed
coordinates $x_{1}$, $x_{2}$, and $x_{3}$. As initial conditions, we
specify a density distribution (atmosphere) and the location of one or
more energy sources and their luminosities (as described in the following
sections). For a summary of these input parameters refer to 
Table~\ref{freepar}.

\subsection{Atmosphere Setup}
\label{atmospar}
We consider two functional forms for the distribution of the ambient
gas. The first form is an exponential density distribution
\begin{equation}
\tilde{\rho} = \exp\left[-\tilde{x}_{2}\right].
\label{rho_exp}
\end{equation}

The second form is the density distribution proposed by
\citet{1990ARA&A..28..215D} (hereafter DL) from their analysis of the
vertical distribution of atomic hydrogen in the Galaxy. This density
distribution, in terms of dimensionless variables, is
\begin{eqnarray}
\label{DLeqn}
\tilde{\rho} = \frac{0.395}{0.566} \exp\left[-\frac{1}{2}\left(a_{1}\tilde{x_{2}}\right)^{2}\right] + \nonumber \\
+ \frac{0.107}{0.566} \exp\left[-\frac{1}{2}\left(a_{2}\tilde{x_{2}}\right)^{2}\right] + \nonumber \\
+\frac{0.064}{0.566} \exp\left[-a_{3}\tilde{|x_{2}|}\right] 
\end{eqnarray}
where $a_{1}=100/90$, $a_{2}=100/225$, and $a_{3}=100/403$. The
coefficients $a_{i}$ express the scale heights of the three components
of the DL layer in units of 100 pc. Contrary to the exponential
profile, the DL layer has an equatorial plane, $\rho(x_{2}=0)=1$, with
density decreasing in both the positive and negative $x_{2}$
directions. This is the same density distribution adopted by
\citet{1998MNRAS.298..797T}.
  
The dynamical importance of the magnetic field is set through the
parameter $\beta_{0}$, defined as the ratio between the gas pressure
and magnetic pressure at $x_{2}=0$
\begin{equation}
\label{beta}
\beta_{0} = \frac{8\pi p}{B^{2}}
\end{equation}

The magnetic field strength is 
\begin{eqnarray}
B = \left(\frac{8\pi p}{\beta_{0}}\right)^{1/2} = \ \ \ \ \ \ \ \ \ \ \ \ \ \ \ \ \ \ \ \nonumber \\
5.0\left(\frac{\rho}{1.67\times 10^{-24}\ \rm g\ cm^{-3}}\right)^{1/2}\beta_{0}^{-1/2}\ \ \mu\rm{G}
\label{Bmagnitude}
\end{eqnarray}
where the numeric expression was derived using $c_{s}^2=\gamma p/\rho$.

We consider two geometries of the magnetic field. The first is a
constant field, i.e. the vertical scale height of the magnetic field
is infinite ($B(x_{2})=B_{0}$), the other takes the scale height of
the magnetic pressure to be equal to the scale height of the gas
pressure ($\beta=\beta_{0}$ everywhere). We refer to the constant
$\beta$ case as equipartition although strictly speaking equipartition
implies $\beta=1$ everywhere. The scale height of the Galactic
magnetic field is not well known. Our initial conditions cover the
possibility of a large scale height $\gtrsim$ 1 kpc corresponding to a
magnetically dominated halo, and a small scale height (equal to 2$H$
for an exponential atmosphere). Hydrostatic equilibrium is assured by
imposing a gravitational potential that balances the gradient of the
total pressure which is the sum of the gas pressure and the magnetic
pressure. The maintenance of hydrostatic equilibrium was tested in
simulations with no source for both atmospheres. The vertical velocity
in these tests was less than $10^{-4}c_{s}$ after 10 Myr simulated
time.  This is consistent with numerical errors and it has no effect
on the results of our simulations.

\subsection{The Energy Source}

The source of the bubble is defined by its mechanical luminosity
($L_{s}$) related to the mass loss rate ($\dot{m}_{s}$) and the
outflow velocity ($v_{s}$) according to
\begin{equation}
L_{s} = \frac{1}{2} \dot{m}_{s}v_{s}^{2}.
\end{equation}

The source is assigned a finite radius ($R_{s}$), determined by the
spatial resolution of the simulation. The mass loss rate from the
source is related to the outflow velocity and the radius of the source
via
\begin{equation} 
\dot{m}_{s} = \rho_{s} v_{s}4\pi R_{s}^{2}.
\end{equation}

Solving for $\rho_{s}$ and converting to dimensionless variables, we find
\begin{equation}
\tilde{\rho_{s}} = \frac{\tilde{L}_{s}}{2\pi\tilde{R_{s}^{2}}\tilde{v_{s}^{3}}}
\label{rhonot}
\end{equation}

\begin{figure}
\centerline{\resizebox{0.6\columnwidth}{!}{\includegraphics[angle=0]{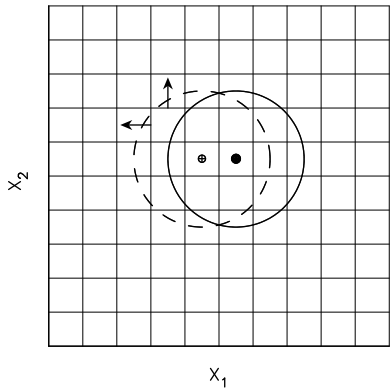}}}
\caption{ The oscillation of the source in $x_1$, $x_2$, and $x_3$
results in a nearly isotropic source with radius only a few zones.
The dashed circle represents the idealized spherical source at one
time, the solid circle represents the idealized source at another
time. Horizontal and vertical lines represent zone boundaries. The two
arrows illustrate velocity vectors for two locations on the a-grid
where mass outflow from the source occurs when the idealized source is
represented by the dashed circle. In the simulations, the amplitude of
the oscillation is set to only half a zone.
\label{grid-fig}
}
\end{figure}

Values for $\tilde{R}_{s}$, $\tilde{L}_{s}$ and $\tilde{v}_{s}$ are
specified as initial conditions. At every time step,
$\tilde{\rho}_{s}$ and $\tilde{v}_{s}$ are reset to their initial
values to maintain constant momentum and energy input from the
source. To simulate a point source, the radius $R_{s}$ should be taken
as small as possible. However, in the rectangular grid, the outflow
from the source will be more isotropic if $R_{s}$ is more than a few
times the grid size. A complication is that density is evaluated in
the center of a pixel (b-grid), whereas velocity is evaluated on the
boundary of a pixel (a-grid). This is part of a strategy to conserve
mass in the simulations, but it introduces an asymmetry in the radius
of a sphere when coordinates are rounded off to integer multiples of
the grid size. To maintain an isotropic source while avoiding
computationally intensive interpolations that must be performed every
time step, the location of the source is oscillated in three
dimensions by half a grid position every time step (see
Figure~\ref{grid-fig}). These excursions of the source along the
three axes are incoherent, and average out most of the anisotropy
introduced by the numerical grid for small $R_{s}$. The isotropy of
the source setup was tested in hydrodynamic simulations with no
density gradient. These experiments showed that $R_{s}$ between two
and three pixels results in good symmetry in the simulations. A small
degree of anisotropy in the outflow of the source remains. This acts
as a seed for the development of instabilities, but is not considered
a problem for our analysis.

\section{RESULTS}
\label{simulations}

\begin{small}
\begin{center}
\begin{deluxetable}{cccccc}
\tablecolumns{6}
\tablewidth{0pt}
\tabletypesize{\scriptsize}
\tablecaption{Physical Parameters for Simulations Performed:\\ Source and Atmosphere \label{simsrun}}
\tablehead{
\colhead{Simulation}& \colhead{$L_{s}$}& \colhead{$n_{0}$ (z = 0)}&  \colhead{B}& \colhead{}&  \colhead{$\beta_{0}$} \\
\colhead{}&  \colhead{($10^{37}$\ \rm erg s$^{-1}$)}& \colhead{(cm$^{-3}$)}&  \colhead{}& \colhead{}& \colhead{}
}
\startdata
\multicolumn{6}{c}{Constant Atmosphere}\\
\hline
ConstH\_zoom& 3& 1&  no B & &$\infty$\\
ConstH & 3& 1 & no B &&$\infty$\\
ConstMHD& 3.2& 0.32&  const & &1.16\\
\cutinhead{Exponential Atmosphere}
ExpH\_zoom& 3& 1& no B  & & $\infty$ \\
ExpH& 3& 1&    no B & &$\infty$ \\
ExpCBa& 3& 1&   const& &10 \\
ExpCBb& 3& 1&  const &&3 \\
ExpCBc& 3& 1&  const &&1 \\
ExpCBd& 3& 1&  const &&0.3 \\
ExpEBa& 3& 1&   equip &&10 \\
ExpEBb& 3& 1&   equip &&3 \\
ExpEBc& 3& 1&   equip &&1 \\
ExpEBd& 3& 1&   equip &&0.3 \\
\cutinhead{DL Atmosphere}
DLH&  3&1&    no B  &&$\infty$ \\
DLCBa& 3& 1&  const &&10 \\
DLCBb& 3& 1&  const &&3 \\
DLCBc& 3& 1&  const &&1 \\
DLCBd& 3& 1&  const &&0.3 \\
DLEBa &3& 1&  equip &&10 \\
DLEBb &3& 1&  equip &&3 \\
DLEBc &3&  1&  equip &&1 \\
DLEBd &3&  1&  equip &&0.3 \\
Tomisaka A &3&  0.3&  const &&0.7 \\
\enddata
\end{deluxetable}
\end{center}
\end{small}

The simulations were performed on the CAPCA\footnote{Computational
Astro-Physics Calgary Alberta (www.capca.ucalgary.ca)} computer
cluster which consists of 64 2.4 GHz Linux based processors connected
by a 1 gigabit network. The results of these simulations were analyzed
using JETGET \citep{2004astro.ph..2121S} and KVIS, a program that is
part of the Karma visualization package \citep{1996ASPC..101...80G}. 

In the simulations that we present here, we examined the effects of
magnetic field strength and geometry on the evolution of a superbubble
in the two atmospheres defined in Section~\ref{atmospar}. The other
parameters were fixed to the values in Table~\ref{simsrun} to apply
our simulations to published results on the bubble associated with the
W4 region as described by \citet{1996Natur.380..687N} and
\citet{1999ApJ...516..843B} (Section~\ref{app}). Table~\ref{simsrun}
gives a summary of the simulations and the values of the varied
physical parameters. Table~\ref{simsrunvolume} gives details on the
simulation volume. To test convergence we ran simulations at
$100^3, 200^3$ and $300^3$ zones and established convergence at
$200^3$ which corresponds to a resolution of 5 pc per voxel (see
Section \ref{sec:x3x1}). 

\begin{small}
\begin{center}
\begin{deluxetable*}{ccccc}
\tablecolumns{5}
\tablewidth{0pt}
\tabletypesize{\scriptsize}
\tablecaption{Box Parameters for Simulations Performed: Volume setup$^{a}$ \label{simsrunvolume}}
\tablehead{
\colhead{Simulation}& \colhead{$x_1\times x_2\times x_3$} & \colhead{$(x_{1} \& x_{3})_{min-max}$} & \colhead{$(x_{2})_{min-max}$}& \colhead{$n_{x_1}\times n_{x_2}\times n_{x_3}$}\\
\colhead{}& \colhead{${\rm pc\times pc\times pc}$}& \colhead{H}&  \colhead{H} &\colhead{${\rm zones\times zones\times zones}$}
}
\startdata
\multicolumn{5}{c}{Constant Atmosphere}\\
\hline
ConstH\_zoom& $300\times 300\times 300$ & $-$1.5 to 1.5 & $-$1 to 2 &  $200\times 200\times 200$\\
ConstH& $10^3\times 10^3\times 10^3$ & $-$5 to 5 & $-$3 to 7 &  $200\times 200\times 200$\\
ConstMHD& $10^3\times 10^3\times 10^3$& $-$5 to 5 & $-$5 to 5 &  $200\times 200\times 200$\\
\cutinhead{Exponential Atmosphere}
ExpH\_zoom& $300\times 300\times 300$& $-$1.5 to 1.5 & $-$1 to 2 &  $200\times 200\times 200$\\
ExpH& $10^3\times 10^3\times 10^3$& $-$5 to 5 & $-$3 to 7 &  $200\times 200\times 200$ \\
ExpCBa& $10^3\times 10^3\times 10^3$& $-$5 to 5 & $-$3 to 7 &  $200\times 200\times 200$\\
ExpCBb& $10^3\times 10^3\times 10^3$& $-$5 to 5 & $-$3 to 7 &  $200\times 200\times 200$\\
ExpCBc& $10^3\times 10^3\times 10^3$& $-$5 to 5 & $-$3 to 7 &  $200\times 200\times 200$\\
ExpCBd& $10^3\times 10^3\times 10^3$& $-$5 to 5 & $-$3 to 7 &  $200\times 200\times 200$\\
ExpEBa& $10^3\times 10^3\times 10^3$& $-$5 to 5 & $-$3 to 7 &  $200\times 200\times 200$\\
ExpEBb& $10^3\times 10^3\times 10^3$& $-$5 to 5 & $-$3 to 7 &  $200\times 200\times 200$\\
ExpEBc& $10^3\times 10^3\times 10^3$& $-$5 to 5 & $-$3 to 7 &  $200\times 200\times 200$\\
ExpEBd& $10^3\times 10^3\times 10^3$& $-$5 to 5 & $-$3 to 7 &  $200\times 200\times 200$\\
\cutinhead{DL Atmosphere}
DLH& $10^3\times 10^3\times 10^3$& $-$5 to 5 & $-$5 to 5 & $200\times 200\times 200$ \\
DLCBa& $10^3\times 10^3\times 10^3$& $-$5 to 5 & $-$5 to 5 & $200\times 200\times 200$\\
DLCBb& $10^3\times 10^3\times 10^3$& $-$5 to 5 & $-$5 to 5 & $200\times 200\times 200$\\
DLCBc& $10^3\times 10^3\times 10^3$& $-$5 to 5 & $-$5 to 5 & $200\times 200\times 200$\\
DLCBd& $10^3\times 10^3\times 10^3$& $-$5 to 5 & $-$5 to 5 & $200\times 200\times 200$\\
DLEBa& $10^3\times 10^3\times 10^3$& $-$5 to 5 & $-$5 to 5 & $200\times 200\times 200$\\
DLEBb& $10^3\times 10^3\times 10^3$& $-$5 to 5 & $-$5 to 5 & $200\times 200\times 200$\\
DLEBc& $10^3\times 10^3\times 10^3$& $-$5 to 5 & $-$5 to 5 & $200\times 200\times 200$\\
DLEBd& $10^3\times 10^3\times 10^3$& $-$5 to 5 & $-$5 to 5 & $200\times 200\times 200$\\
Tomisaka A& $10^3\times (2\times10^3) \times 10^3$&0 to 10 & $-$10 to 10 & $200\times 400\times 200$ \\
\enddata
\tablenotetext{a}{This table gives information on the main set of simulations. Additional simulations
for convergence tests with $100\times 100\times 100$ and $300\times 300\times 300$ zones 
(see Figure~\ref{fig:x3x1conv}) are not listed. }
\end{deluxetable*}
\end{center}
\end{small}

\subsection{Comparison to Analytical Solutions}

\subsubsection{Hydrodynamic Solutions}

\begin{figure*}
\centerline{\resizebox{0.9\textwidth}{!}{\includegraphics[angle=-90]{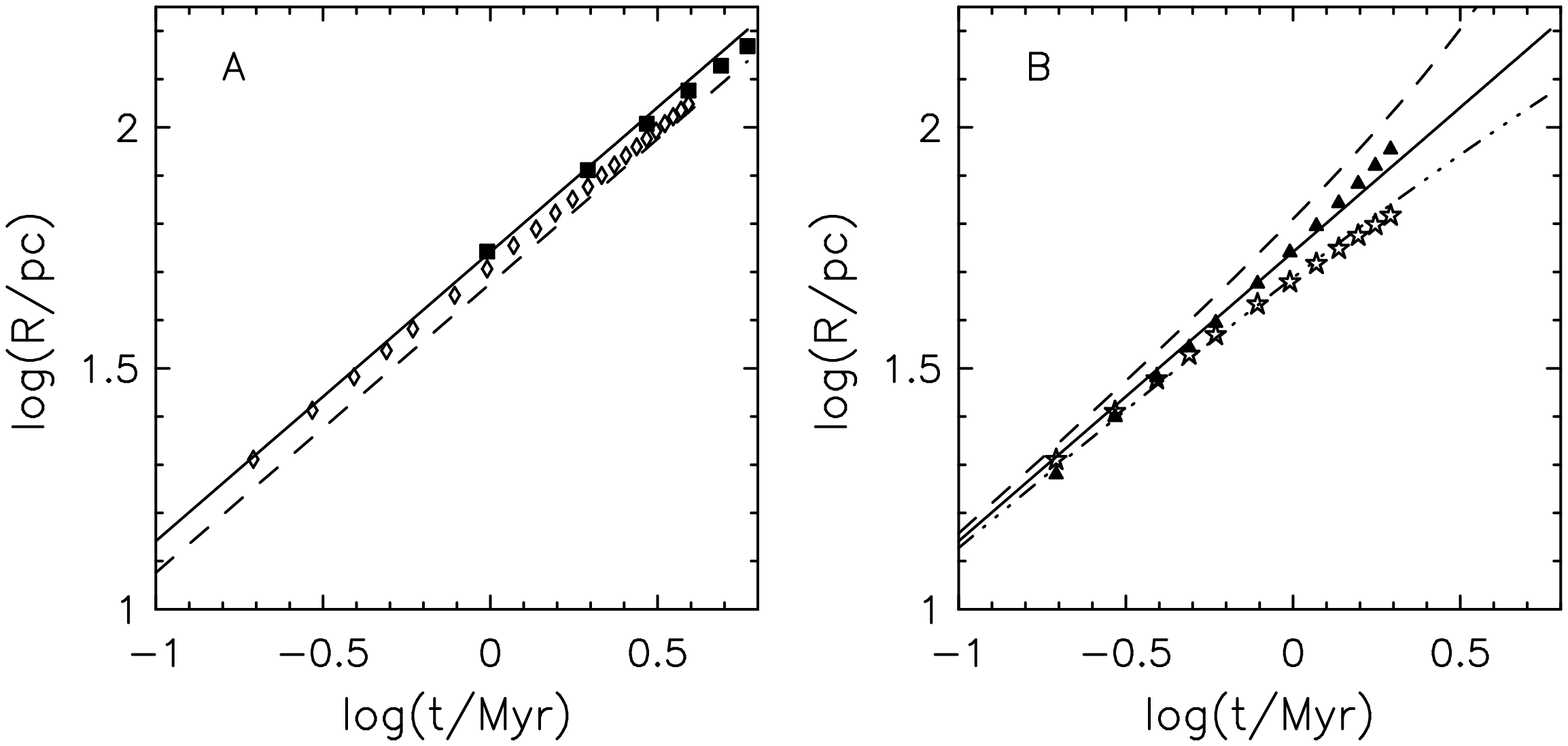}}}
\caption{\textbf{(A)} Time evolution of the radius of a stellar wind
bubble in a medium with constant density and no magnetic
field. Diamonds show the radius of the bubble in a simulation with
$200 \times 200 \times 200$ zones that measures $300 \times 300 \times
300$ pc (i.e. a resolution of 1.5 pc per voxel). Squares show the
radius of a bubble in a $200 \times 200 \times 200$ simulation that
measures $1000 \times 1000 \times 1000$ pc (i.e. a resolution of 5 pc
per voxel). In both cases the radius of the source was 2 pixels. The
radii of the simulations are compared with the radius of the contact
discontinuity in the \citet{1977ApJ...218..377W} model (dashed
line). Also shown is the radius of the outer shock in the
\citet{1977ApJ...218..377W} model (solid line). \textbf{(B)} Time
evolution of the radius of the bubble in an exponential atmosphere in
the positive $x_{2}$ direction (filled triangles) and the negative
$x_{2}$ direction (stars). The corresponding radii from the K60 model
are shown as a dashed curve and a dot-dashed curve respectively. Also
shown is the \citet{1977ApJ...218..377W} constant density model from
panel A (solid line). See Appendix A for a discussion on the
difference between the simulations and the K60 model at early times.
\label{radiusgraph}
} 
\end{figure*}

\label{comparetoanalytic}
In order to test the setup, we performed hydrodynamic simulations
with a constant density profile (ConstH\_zoom \& ConstH) to compare
with the \citet{1975ApJ...200L.107C} model and a hydrodynamic
simulation with an exponential density profile (ExpH\_zoom) to compare
with the K60 model. The physical parameters for
these simulations are given in Table~\ref{simsrun}. Maintaining the
same number of pixels for a smaller volume (300 pc on a side) results
in a finer resolution (1.5 pc compared to the 5 pc resolution of the
other simulations listed in Table~\ref{simsrunvolume}). This allows us
to follow the evolution of the bubble in detail at early times. We
have multiple reasons to consider high-resolution simulations for the
comparison with the \citet{1975ApJ...200L.107C} and
K60 models. The first reason is to investigate the
effect of resolution on our solutions. The second reason is to explore
the effect of the finite source size. In our simulations, the bubble
begins at a finite radius at $t$ = 0 (the radius of the source) which
can be made smaller in high resolution simulations.  Although the
\citet{1975ApJ...200L.107C} model is self-similar, increasing the
resolution will allow us to investigate the effect of the finite
source size in the simulations at early times. The
K60 model is not self-similar, but it is only
expected to agree with the simulations at early times
(Section~\ref{analytic}).
   
Both models assume that the sound speed in the ambient medium is zero
by assuming the pressure in the undisturbed medium is negligible. For
our simulations, the sound speed in the undisturbed medium is
finite, but the assumption of negligible pressure is satisfied as
long as the pressure inside the bubble is much larger than the
pressure of the surrounding atmosphere. The limit of negligible
ambient pressure adopted in the K60 model implies
$t_{o}\gg t_{o_{k}}$. In our simulations, the time scale $t_{o}$
(Equation~\ref{dimlesstime}) is $\sim10^{7}$ years, whereas the
timescale $t_{o_{k}}$ of the K60 model
(Equation~\ref{ytilde}) is $\sim0.05\times10^{7}$ years. The
simulations should therefore resemble the Kompaneets model at
times much less than 10$^{7}$ years.

\begin{figure*}
\centerline{\resizebox{0.7\textwidth}{!}{\includegraphics[angle=0]{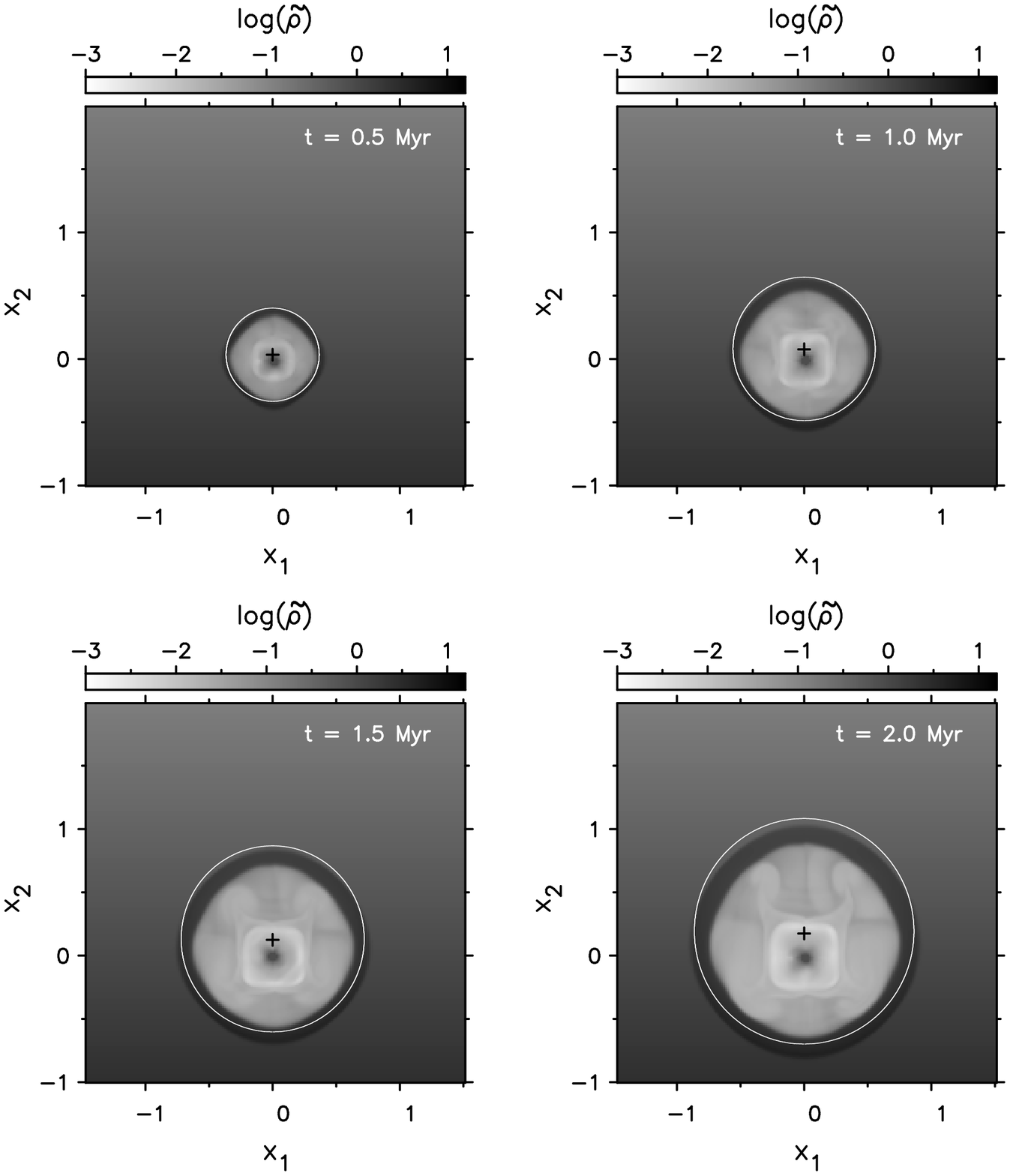}}}
\caption{Early stages of a zoomed-in hydrodynamic simulation with
Kompaneets solution overlaid (white circles) in a plane through the
centre of the source at ($x_{1}$,$x_{2}$,$x_{3}$)=(0,0,0). Gray scales
represent density on a logarithmic scale from 10$^{-3}$ to 10$^{1.2}$
cm$^{-3}$. At early times, the Kompaneets model is nearly spherical,
but the center of the sphere has a visible offset in the direction of
the density gradient (see Appendix~A). The geometric center of the
Kompaneets solution as defined in Appendix A is shown as a $+$ in each
panel. The values for $\tilde{y}$ (Equation~\ref{ybasu}) are 0.36,
0.55, 0.70, and 0.84 for these panels in increasing time order. The
unit of length on the axes is 100 pc.
\label{kompaneetsfig}
} 
\end{figure*}

The size of the bubble was parameterized by the radius of the contact
discontinuity, which is well defined in both the simulations and the
Kompaneets model. The shell of swept-up interstellar medium is presumed
to be infinitely thin in the Kompaneets model, so the radius of the
outer shock is not predicted. For consistency, the same was done for
the spherical case. The radius of the contact discontinuity for the
spherical model was taken to be 86\% of the outer shock radius
following \citet{1977ApJ...218..377W}. For ConstH\_zoom, the contact
discontinuity is located at 84\% of the outer shock radius, which is
in good agreement with the \citet{1977ApJ...218..377W}
model. Figure~\ref{radiusgraph}A compares the time evolution of the
radius of ConstH\_zoom with the spherical model. The simulation
displays a power law expansion with time $R_{\rm b}\sim t^{\alpha}$
with $\alpha=0.566$.  This is only 5.7\% smaller than the slope of the
\citet{1975ApJ...200L.107C} model, which has $\alpha=0.6$. At an age
of 1 Myr the radius of ConstH is 9\% larger than the radius of the
model. The filled squares in Figure~\ref{radiusgraph}A show the time
evolution of ConstH. The radius of the bubble in this simulation is at
most 10\% larger than the radius in the higher resolution simulation
at the same age. This indicates that part of the difference between
our simulations and the Castor model may be related to the finite size of
the source.

Figure~\ref{radiusgraph}B shows the time evolution of the
radius\footnote{Here, the radius is defined separately for each
direction as the distance from the source to the contact discontinuity
at the top of the bubble and the distance from the source to the
contact discontinuity at the bottom of the bubble.} of a bubble in an
exponential atmosphere (ExpH$_{-}$zoom in Tables~\ref{simsrun} and
\ref{simsrunvolume}), looking perpendicular to the Galactic plane, for
the simulations and the K60 model. Also shown is the spherical model
from Figure~\ref{radiusgraph}A as a reference line. The bottom of the
simulation closely follows the K60 model. However the top of the K60
model is consistently higher than the simulation by about 16\% even at
early times. This is also shown in Figure~\ref{kompaneetsfig}. As time
proceeds, the difference between the K60 model and the simulations
increases significantly at the top of the bubble but not at the
bottom. \citet{komljenovic1999} also compared their 2-dimensional
hydrodynamic simulations to the Kompaneets model with similar
results. \citet{MacLow1989} found consistency between their Kompaneets
approximation from \citet{MaclowandMcCray1988} and hydrodynamic
simulations at times as late as 6.87 Myr. However, their Kompaneets
approximation is a solution in an atmosphere with an equatorial plane
unlike the Kompaneets model \citep{Kompaneets1960} considered here.
The axial ratio in the Galactic plane (i.e. $x_{1}x_{3}$) is expected
to be unity at all times for the hydrodynamic case and also at early
times in the MHD case (see Section~\ref{analysis}). The radius of the
cross-section through the source of the K60 model as a function of
time can be calculated to a good approximation from the spherical
Castor model, as shown numerically by \citet{1999ApJ...516..843B}.

\subsubsection{MHD Analytical Approximation}

\citet{Ferriere1991} derived analytic solutions of a superbubble in a
uniform magnetic field in the limit of high expansion velocity, and
numerical solutions for the general case of smaller expansion
velocity. To compare with their solutions, we ran an MHD simulation in
an atmosphere with constant density (ConstMHD, see
Tables~\ref{simsrun} \& \ref{simsrunvolume}). \citet{Ferriere1991}
found that the outer shock front remains nearly spherical in a uniform
magnetized medium, but the cavity is smaller in the direction
perpendicular to the magnetic field, compared with the hydrodynamic
solution. We find the same in our numerical simulation, but we see no
dimple in the outer shock in the direction of the magnetic
field. \citet{1998MNRAS.298..797T} also did not see this dimple. The
expansion along the magnetic field lines is nearly the same as in the
hydrodynamic simulation as found by \citet{Ferriere1991}.

The thickness of the shell perpendicular to the magnetic field
in the analytic approximation found by \citet{Ferriere1991}, increases linearly
with time according to
\begin{equation}
\Delta R = 1.2 \frac{B_{-6}t_{6}}{\sqrt{n_{0}}} \sin \theta\ \rm pc
\label{ferriere}
\end{equation}
where $n_{0}$ is the number density in cm$^{-3}$, $B_{-6}$ is the
ambient magnetic field strength in $\mu$G, and $t_{6}$ is the time in
Myr. Our simulation reproduces the linear increase of shell thickness
perpendicular to the magnetic field with time, but the shell thickness
is a factor $\sim$ 2.6 larger. The simulations by
\citet{1998MNRAS.298..797T} also show a very thick shell perpendicular
to the magnetic field, although he did not consider a medium with
constant density. The reason for the difference is that
Equation~(\ref{ferriere}) was derived in the limit of high expansion
velocity, defined as a small ratio of magnetic pressure to ram
pressure. This ratio increases with time as the outer shock
decelerates. We find that this ratio is of order unity or larger at
times $\gtrsim$ 2.5 Myr in our simulation. Simulations with higher
$\beta$ reproduce the shell thickness in Equation~(\ref{ferriere}),
while also satisfying the assumptions made in the derivation of this
equation.

\subsection{Exponential Atmosphere}
\label{Exp}
\subsubsection{Model ExpH}
\label{ExpH}
A simulation of a bubble expanding in an exponential atmosphere with
no magnetic field was done primarily to compare with previously
published results and to obtain the limiting case $\beta=\infty$. At
early times, the bubble remains spherical
(Figure~\ref{kompaneetsfig}).  The axial ratio of the cavity in the
simulations {\it and the axial ratio of the Kompaneets model} are
indeed found to be very close to unity, even though
Figure~\ref{kompaneetsfig} shows that the top of the Kompaneets model
has proceeded significantly further than the simulation even at these
early times. In Appendix~A we show that the Kompaneets model at early
times can be described to third order in $\tilde{y}$ as a spherical
cavity that rises in the atmosphere. Although the axial ratio remains
unity to a high degree of accuracy, the geometric center of the model,
defined by the point midway between the top and the bottom (crosses in
Figure~\ref{kompaneetsfig}), is displaced significantly from the
location of the source.

At later times, the bubble grows larger than the scale height of the
medium and accelerates in the vertical direction. As the bubble
expands into the upper atmosphere, the top of the bubble continues to
accelerate and a Rayleigh-Taylor instability develops when the top
reaches $\sim 4$ scale heights above the source, which occurs at an
age of $\sim$10 Myr. The bubble breaks out of the galactic plane to
form a chimney after about 16 to 17 Myr.

\subsubsection{Models ExpCB(a-d) \& ExpEB(a-d)}
\label{ExpCBandE}

\begin{figure*}
\resizebox{\textwidth}{!}{\includegraphics[angle=-90]{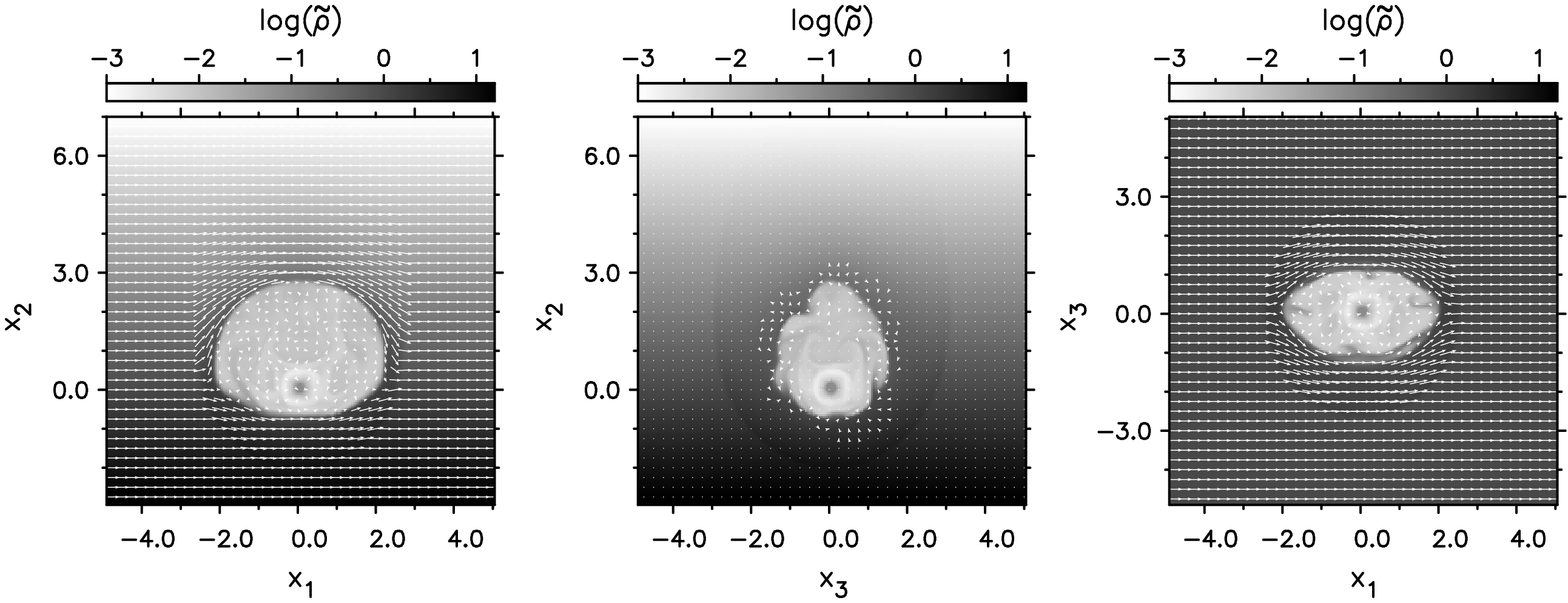}}
\caption{Simulation ExpCBc at an age of 10 Myr. Magnetic field
initially oriented along x$_{1}$ axis, $\beta = 1$. Panels show slices
through the cube at the location of the source,
($x_{1}$,$x_{2}$,$x_{3}$)=(0,0,0), in three orthogonal
planes. Grayscales show the gas density on a logarithmic scale from
$10^{-3}$ (white) to $10^{1.2}$ (black) cm$^{-3}$. The vector field
depicts the projection of magnetic field vectors on the plane of the
image. The unit of length on the axes is 100 pc.
\label{ExpConst_B_40}
} 
\end{figure*}

\begin{figure*}
\resizebox{\textwidth}{!}{\includegraphics[angle=-90]{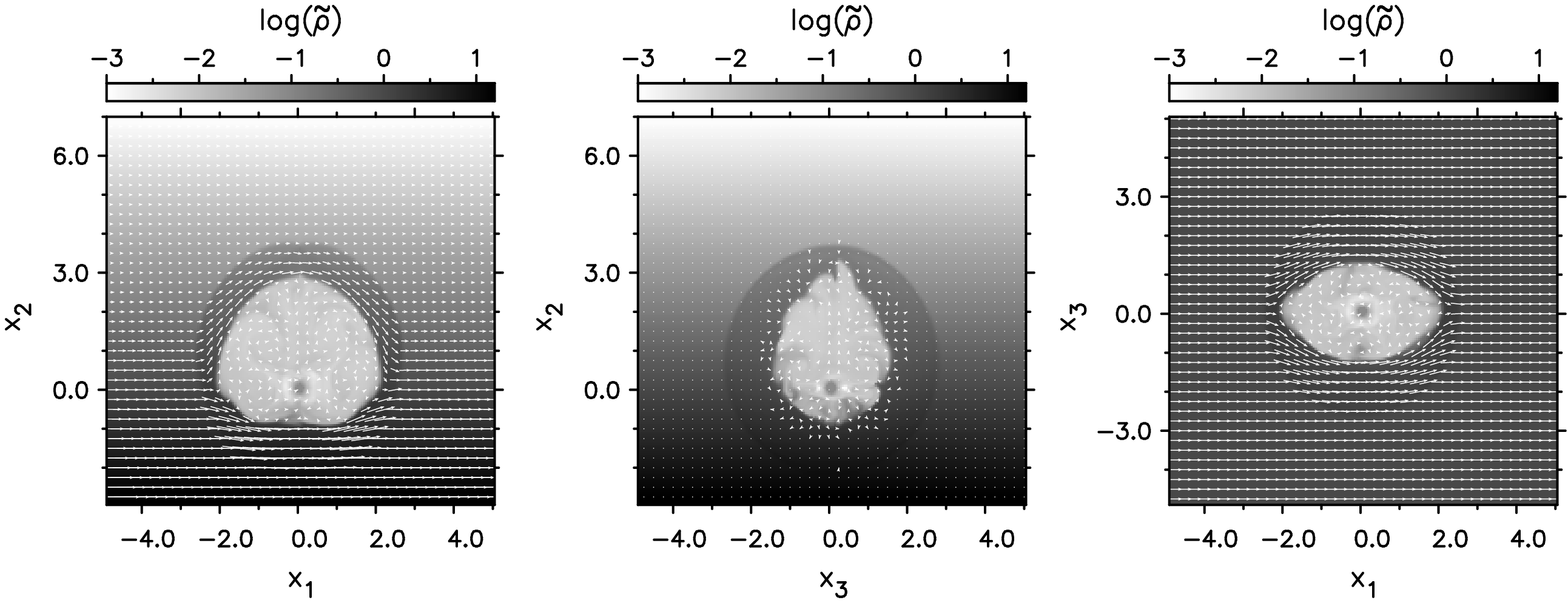}}
\caption{Same as Figure \ref{ExpConst_B_40} but for Simulation ExpEBc at an age of 10 Myr. 
\label{ExpEquip_B_40}
} 
\end{figure*}

The main difference between a constant magnetic field and
equipartition magnetic field is in the evolution of the bubble at
large distances from the Galactic plane. A constant magnetic field
implies a high Alfv\'en speed in the Galactic halo whereas constant
$\beta$ implies that the Alfv\'en speed in the halo is the same as in
the disk. This difference has significant consequences for the
structure of the shock as it expands from the disk into the
halo. Figure~\ref{ExpConst_B_40} shows a bubble in an exponential
atmosphere with constant magnetic field ($\beta_{0} = 1$) oriented
along the $x_{1}$ axis, at an age of 10 Myr.  

The cavity is elongated along the magnetic field as the plasma can
move freely along the field lines but its motion perpendicular to the
field lines is restricted. However, the shape of the outer shock at
the level of the source in the Galactic plane ($x_{1}x_{3}$ plane) is
nearly circular because the propagation speed of the outer shock is
similar in both directions. Along the $x_{1}$ axis, the outer shock
and the contact discontinuity are close together resulting in a thin
compressed shell of swept-up interstellar medium. Along the $x_{3}$
axis, perpendicular to the initial magnetic field, the distance
between the outer shock and the contact discontinuity is much larger
resulting in a thick shell. Looking along the field lines, the cavity
is more elongated than in the hydrodynamic simulation, and the shell
of swept-up interstellar medium is thicker than in the hydrodynamic
case. At the top of the cavity a fast magnetosonic wave runs upwards
into the halo, with very little compression of the halo gas. In
contrast to the hydrodynamic simulation, we see no evidence for a
Rayleigh-Taylor instability developing at the top of the bubble,
because the top of the bubble does not accelerate. Instead, an
instability develops everywhere at the contact discontinuity as is
apparent from the scalloped shape of the contact discontinuity in the
$x_{3}x_{2}$ plane (looking along the field lines). This instability
also occurs in the simulations of \citet{1998MNRAS.298..797T} and may
be similar to the magnetically enhanced Rayleigh-Taylor instability
proposed by \citet{gregori2000}. At the time shown in
Figure~\ref{ExpConst_B_40}, the expansion of the cavity along the
$x_{3}$ axis (perpendicular to the magnetic field) in the Galactic
plane has stalled as a result of magnetic tension. The cavity still
expands along the $x_{1}$ and $x_{2}$ axes.

Figure~\ref{ExpEquip_B_40} shows simulation ExpEBc also at the age of
10 Myr. In contrast with Figure~\ref{ExpConst_B_40},
Figure~\ref{ExpEquip_B_40} shows a shell of compressed interstellar
medium at the top of the bubble. Since the Alfv\'en velocity does not
increase with distance from the Galactic plane, the outer magnetosonic
shock does not accelerate into the halo but remains close to the
contact discontinuity. The shell appears approximately twice as thick
looking along the field lines as looking perpendicular to the magnetic
field. The axial ratio of the cavity in the $x_{3}x_{2}$ plane
(looking along the field lines) is smaller than in the ExpCBc
simulation. This indicates the importance of magnetic tension
confining the bubble, even though the total pressure mimics the
pressure in the hydrodynamical exponential atmosphere. The magnetic
field in the upper part of the shell is significantly enhanced
compared to the magnetic field in the undisturbed atmosphere at the
same height above the plane. Here too, we see an instability
everywhere along the contact discontinuity in the $x_{3}x_{2}$ plane.

\subsection{Dickey \& Lockman Atmosphere}
\label{DLM}
\subsubsection{Model DLH}
\label{DLH}

\begin{figure}
\resizebox{\columnwidth}{!}{\includegraphics{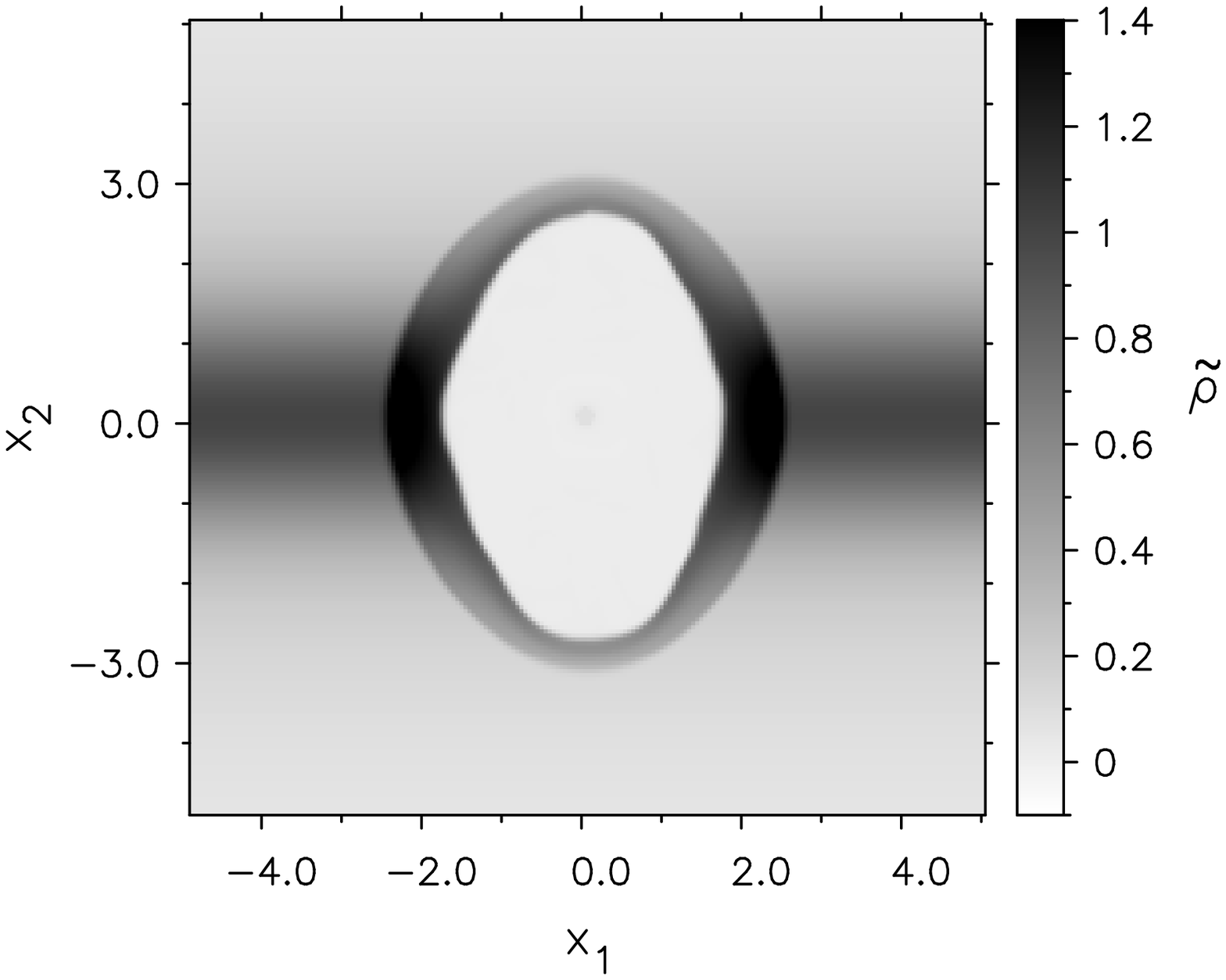}}
\caption{Simulation DLH (hydrodynamic) at an age of 10 Myr. The
density in a slice through the cube at the location of the source,
($x_{1}$,$x_{2}$,$x_{3}$)=(0,0,0), in the $x_{1}x_{2}$ plane is shown
with linear grayscales from $-0.2$ (white) to 1.4 (black)
cm$^{-3}$. The unit of length on the axes is 100 pc.
\label{fig:DLH}
}
\end{figure}

The simulations in the DL atmosphere show some differences compared to
the evolution in the exponential atmosphere. The equatorial plane
introduces symmetry with respect to the plane $x_{2}=0$. In the
absence of a magnetic field the bubble is spherical until it expands
to radius $\sim 200$ pc. Once the radius of the bubble grows beyond
$\sim$200 pc, it becomes more elongated in the vertical direction as it
balloons out into the halo (Figure~\ref{fig:DLH}).  The top and bottom of the
bubble expanding into the low-density outer atmosphere, accelerate away
from the Galactic plane, as in the case of an exponential atmosphere.
The onset of a Rayleigh-Taylor instability is not seen in this
simulation but it may develop at later times outside the simulation
volume.

Between an age of $\sim14$ Myr and $\sim18$ Myr a small decrease in
the radius of the cavity in the equatorial plane is observed. A
similar stall of the expansion and subsequent contraction in the
Galactic plane triggered by the rapid vertical expansion of the bubble
was observed by \citet{1998MNRAS.298..797T}.

\subsubsection{Models DLCB(a-d) \& DLEB(a-d)}
\label{DLCBandE}

\begin{figure*}
\resizebox{\textwidth}{!}{\includegraphics[angle=-90]{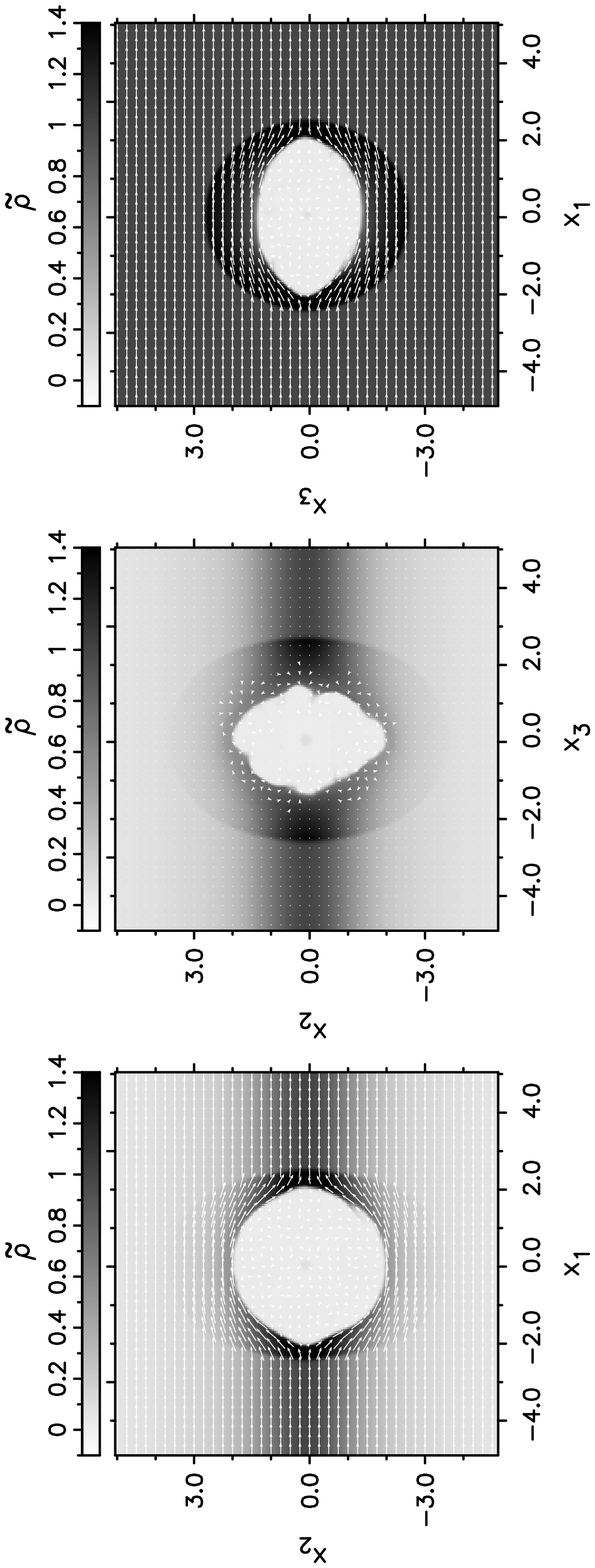}}
\caption{Simulation DLCBc at an age of 10 Myr. Panels show slices
through the cube at the location of the source,
($x_{1}$,$x_{2}$,$x_{3}$)=(0,0,0), in three orthogonal
planes. Grayscales show the gas density on a linear scale from $-0.2$
(white) to 1.4 (black) cm$^{-3}$. The vector field depicts the
projection of magnetic field vectors on the plane of the image. The
unit of length on the axes is 100 pc.
\label{B40fig}
} 
\end{figure*}

\begin{figure*}
\resizebox{\textwidth}{!}{\includegraphics[angle=-90]{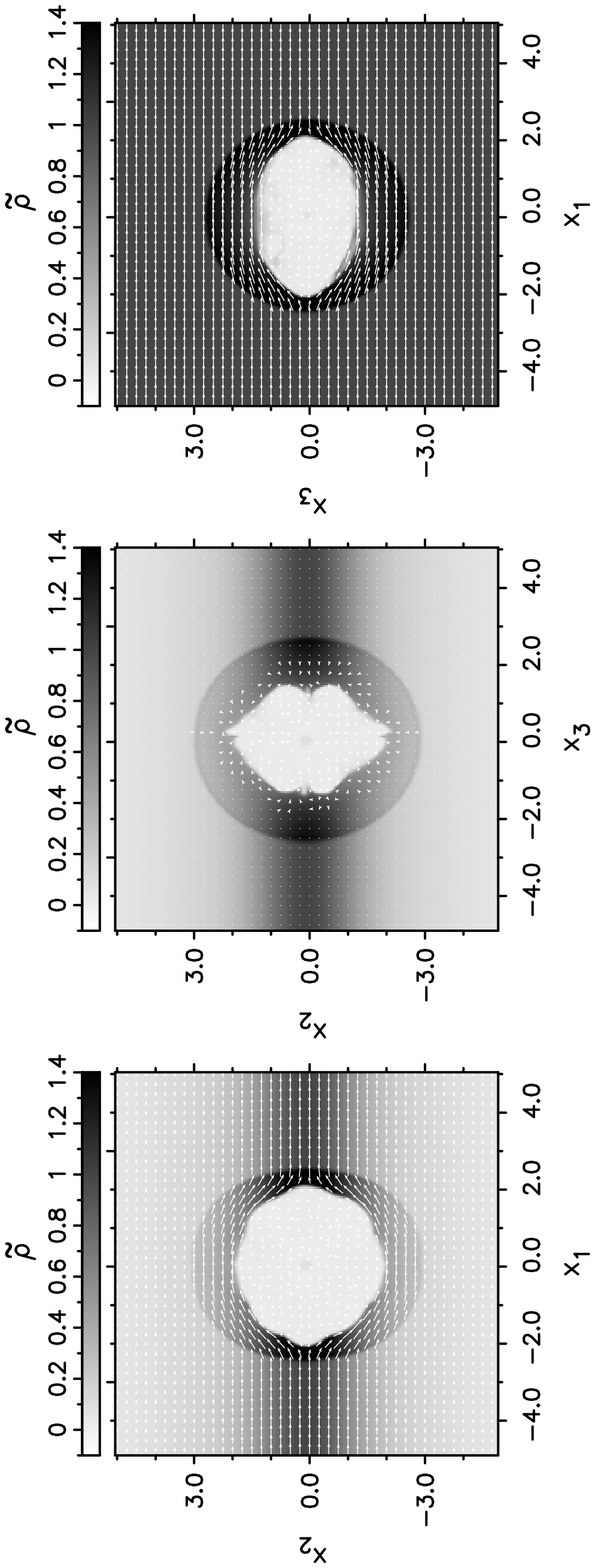}}
\caption{Same as Figure \ref{B40fig} but for Simulation DLEBc at an age of 10 Myr.  
\label{DLEquip_B_40}
} 
\end{figure*}

Figure~\ref{B40fig} shows the simulation of a bubble in the DL
atmosphere with constant magnetic field oriented along the $x_{1}$
axis at an age of 10 Myr. As with the exponential simulations, this
simulation shows the same features as discussed by
\citet{1998MNRAS.298..797T}. The cavity is elongated along the
magnetic field lines. The shape of the outer shock at the level of the
source in the $x_{1}x_{3}$ plane is nearly circular, similar to the
exponential case. This is to be expected because the density and
magnetic field strengths are the same at the level of the source in
both atmospheric profiles. In contrast to the exponential atmosphere,
the shape of the cavity in the $x_{1}x_{2}$ plane is fairly circular
even though the bubble expands to a size much larger than the pressure
scale height. The shape of the cavity in the $x_{3}x_{2}$ plane is
quite elongated and displays the same magnetically enhanced
Rayleigh-Taylor instability as discussed in Section~\ref{ExpCBandE}.
As with the ExpCBc case, a fast magnetosonic wave runs upwards into
the halo. Figure~\ref{DLEquip_B_40} shows the constant $\beta$
simulation DLEBc at the same age. The discussion of the differences
between the ExpCBc and ExpEBc applies also to simulations in the DL
atmosphere.

We performed an extra simulation (Tomisaka A) in a larger volume to
compare with Model A in \citet{1998MNRAS.298..797T}, with the same
equatorial density. Other initial conditions were set as closely as
possible to those specified by \citet{1998MNRAS.298..797T} (see
Tables~\ref{simsrun}~\&~\ref{simsrunvolume}), but some small differences in
the setup could not be avoided. Our simulations show the same
characteristics in terms of the size and shape of the cavity, and the
thickness of the surrounding shell. However we found differences of
the order of 20\% in the dimensions of the cavity after 10 Myr which
appear to be associated with small differences in initial
conditions. The fast magnetosonic shock had proceeded to almost twice
the height of the shock in Model A of \citet{1998MNRAS.298..797T},
indicating a lower value of $\beta$ in our simulation. The shape of
the magnetosonic shock front is particularly sensitive to small
differences in $\beta$ because the Alfv\'en speed at high altitudes is
very sensitive to $\beta$.

\vfill
\subsection{Bubble Morphology as a Function of Time and Magnetic Field Strength}
\label{analysis} 

Our simulations cover a factor 30 in magnetic field strength, from
relatively weak (those labeled with 'a', $\beta = 10$) to strong
(those labeled with 'd', $\beta = 0.3$). Here we discuss variation in
the shape of a bubble with magnetic field strength at a reference age
of 10 Myr. The weak magnetic field limit for the simulations converges
to the hydrodynamic case as should be expected. The differences
between simulations with $\beta = 10$ and hydrodynamic simulations are
minor. We limit the discussion to the range $\beta = 10$ to 0.3.

\begin{figure*}
\resizebox{0.5\textwidth}{!}{\includegraphics[angle=0]{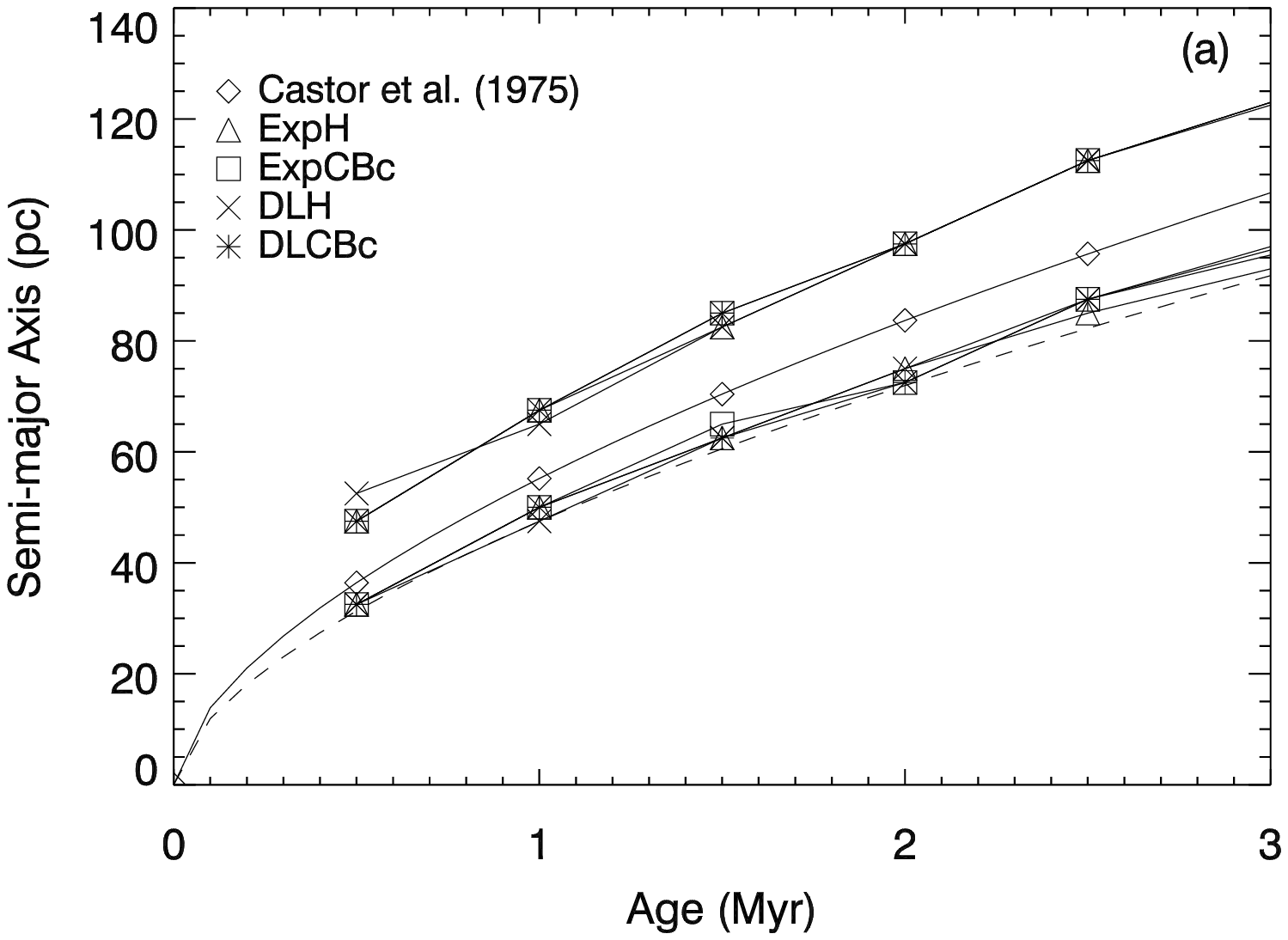}}
\resizebox{0.5\textwidth}{!}{\includegraphics[angle=0]{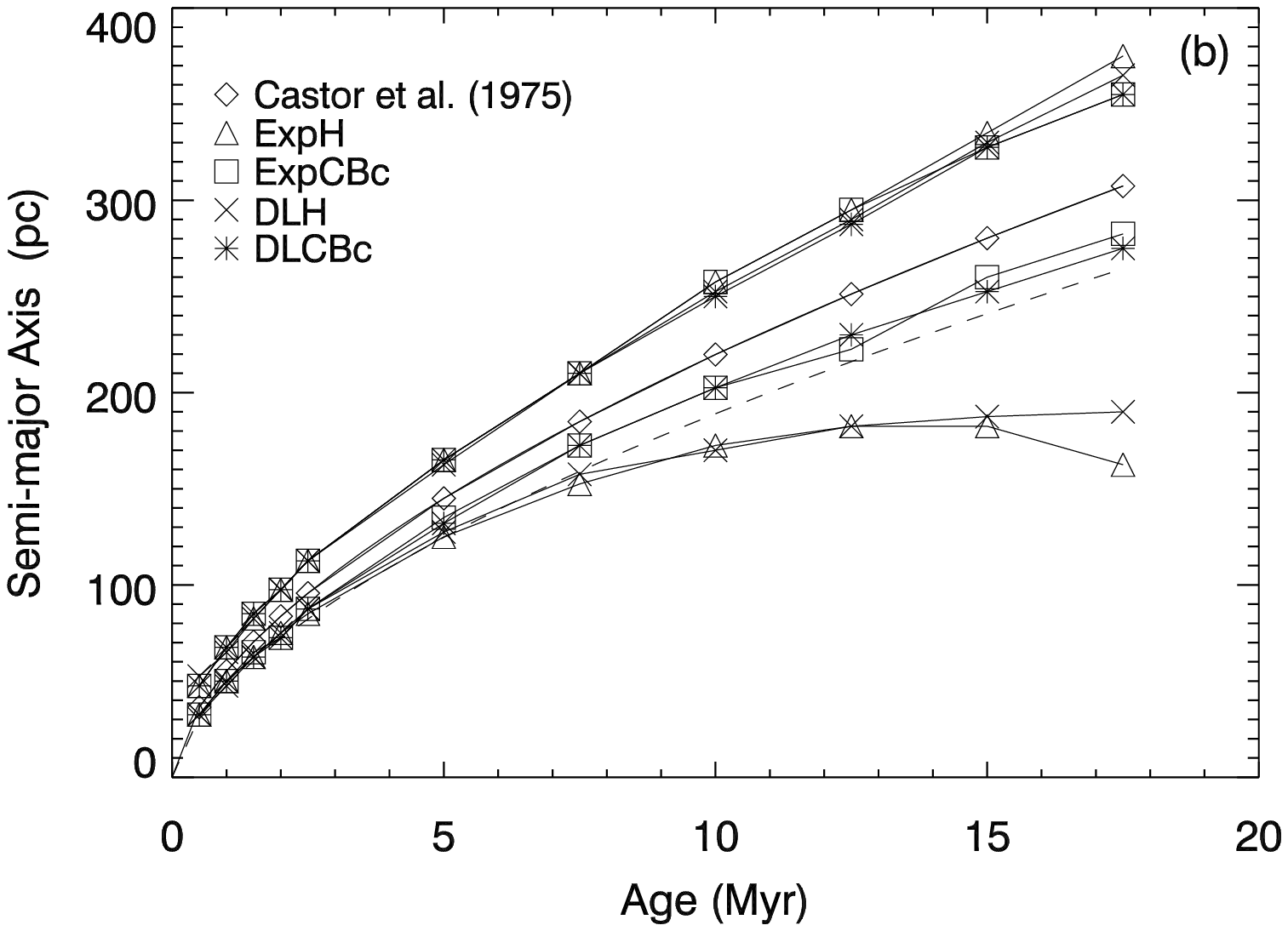}}
\caption{Comparison of the semi-major axis in the $x_1 x_3$ plane of
the outer shock and contact discontinuity, for a representative set of
simulations, to the \citet{1975ApJ...200L.107C} model. The semi-major
axis is defined as half the diameter in the $x_1$ direction at the
level of the source. Diamonds give the radius of the outer shock in
the \citet{1975ApJ...200L.107C} model, where the contact discontinuity
is located at 86\% of the radius of the outer shock (indicated by the
dashed curve).  (a): Enlargement of the evolution at early times. (b):
Evolution at later times.}
\label{fig:castorcompare}
\end{figure*}

One of the most visible effects of the magnetic field is on the size
of the cavity. A decrease in $\beta$ by a factor of 30 results in a
decrease in the vertical size of the cavity by a factor $\sim$1.7,
independent of the density and magnetic field stratifications.  The
diameter of the cavity in the $x_1$ direction (along the field lines)
increases by a factor of $\sim$ 1.4 as $\beta$ decreases. Expansion
along the magnetic field is faster as the magnetic field is stronger.
The diameter of the cavity along the $x_3$ axis decreases by a factor
$\sim$1.6 as $\beta$ decreases from 10 to 0.3. The effect of a
stronger magnetic field on the shape of the cavity is that the cavity
becomes more elongated along the Galactic plane when observed along a
direction perpendicular to the Galactic magnetic field. When observed
along the direction of the magnetic field, the shape of the cavity
varies less, but the size of the cavity becomes significantly smaller.

The thickness of the shell in the directions perpendicular to the
magnetic field also increases dramatically with increasing magnetic
field strength. However, this happens at the expense of the amount of
compression of gas in the shell, which becomes nearly invisible in the
simulations with $\beta = 0.3$. In practice, the shell may not be
detectable in real data confused with emission in the foreground and
background, even for $\beta = 1$, if the magnetic field scale height
is large (see Figure~\ref{ExpConst_B_40}).  The increasing thickness
of the shell along the $x_3$ axis is partly the result of the smaller
extent of the cavity, partly because of a faster expansion of the
outer shock, approximately by equal amounts.  In simulations with a
constant magnetic field, the thickness of the shell at the top of the
bubble increases mainly because of the fast upward expansion of the
outer shock, and to a lesser amount by the smaller extent of the
cavity. In the equipartition simulations, the increased thickness of
the top of the shell is mainly because of the smaller extent of the
cavity. Along the magnetic field lines ($x_1$), the thickness of the
shell decreases as the magnetic field strength increases. However, no
strong increase in the density is found at the extremes of the shell
along the $x_1$ axis.

Figure~\ref{fig:castorcompare} compares the time evolution of the
semi-major axis of the contact discontinuity and the outer shock in
the $x_1x_3$ plane, for a sample of our simulations (ExpH, DLH, ExpCBc, \&
DLCBc), with that of the \citet{1975ApJ...200L.107C} model.  The
semi-major axis is defined as half the diameter along the $x_1$ axis
at the level of the source. The expansion along the magnetic field
lines at early times (see Figure~\ref{fig:castorcompare}a) closely
follows the hydrodynamic solution in accordance with
\citet{Ferriere1991}. The contact discontinuity of the self-similar
solution was found to be at 86\% of the shock radius by
\citet{1977ApJ...218..377W}. The radius of the contact discontinuity
in the analytic model indeed agrees well with the radii of the contact
discontinuity in our simulations at early times (dashed curve in
Figure~\ref{fig:castorcompare}). At later times
(Figure~\ref{fig:castorcompare}b), the contact discontinuity in the
ExpCBc and DLCBc simulations has proceeded further than the analytic
solution (dashed curve). On the other hand, the contact discontinuity
in the ExpH and DLH simulations lags behind the analytic solution. We
interpret this behavior as the result of the hydrodynamic simulations
breaking out of the disk, reducing the pressure inside the cavity,
whereas the MHD simulations remain confined to the disk. We find that
the location of the outer shock in these simulations is nearly the
same even after 15 Myr in agreement with results presented in
\citet{Ferriere1991}.

\subsubsection{Axial Ratios}\label{sec:x3x1}

\begin{figure}
\resizebox{\columnwidth}{!}{\includegraphics[angle=0]{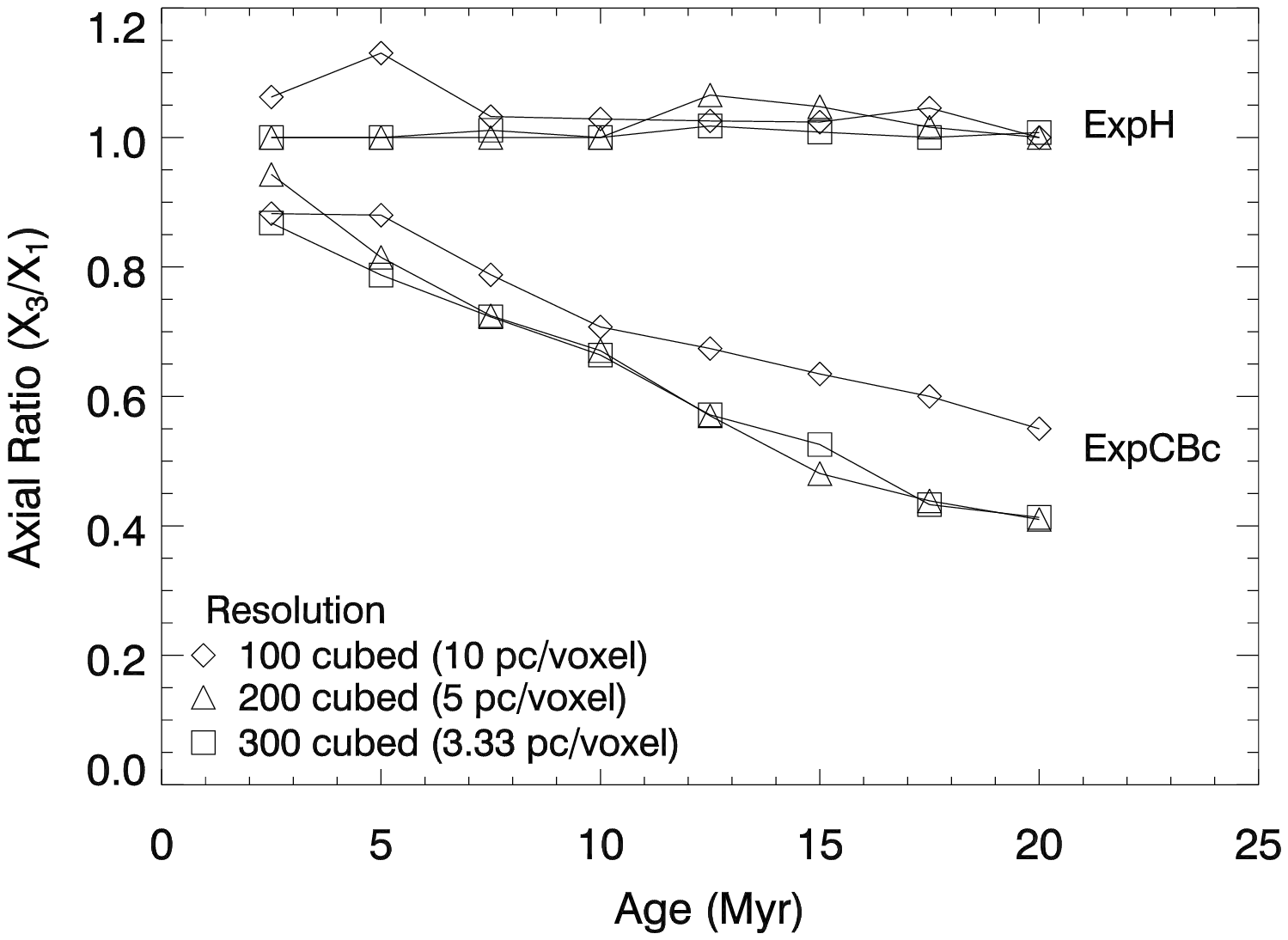}}
\caption{Axial ratios ($x_{3}$/$x_{1}$) of simulated bubbles as a
function of time for different resolutions indicating convergence at
$200 \times 200 \times 200$ zones.
\label{fig:x3x1conv}
} 
\end{figure}

The discussion in the previous subsection shows that the magnetic
field is as important as the density distribution of the medium in
determining the shape of a superbubble. Observationally, the shape of
the cavity is most important because structures with a large contrast
in density are most easily detected in \HI\ images of the Galaxy.

\begin{figure}
\resizebox{\columnwidth}{!}{\includegraphics[angle=0]{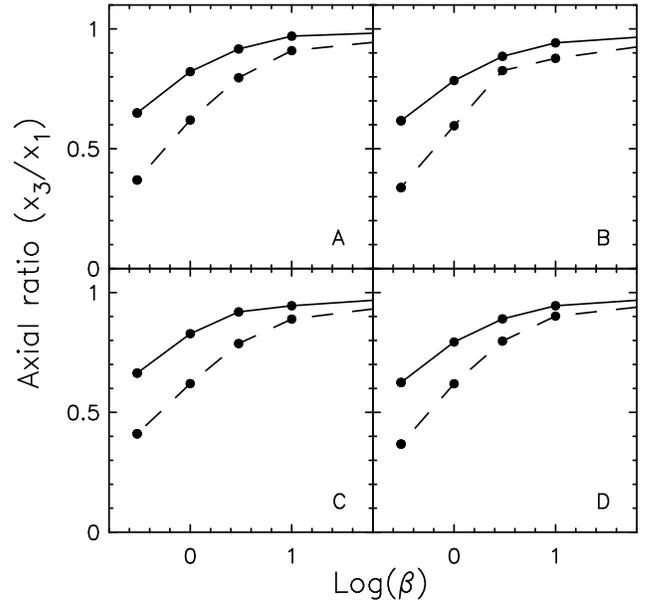}}
\caption{Axial ratios ($x_{3}$/$x_{1}$) of simulated bubbles as a
function of $\beta$ at an age of 5 Myr (solid curve) and 12.5 Myr
(dashed curve). In each panel dots represent the axial ratio measured
in our simulations. The curves in all four panels converge to an axial
ratio of unity in the limit $\beta \rightarrow \infty$ (the
hydrodynamic case). (A) Exponential atmosphere with constant magnetic
field (Simulations ExpCBa-d), (B) Exponential atmosphere with
constant $\beta$ (ExpEBa-d), (C) Dickey and Lockman atmosphere with
constant magnetic field (DLCBa-d), (D) Dickey and Lockman atmosphere
with constant $\beta$ (DLEBa-d).  The axial ratio at the beginning of
the simulations is unity for each $\beta$.
\label{x1x3ratio}
} 
\end{figure}

Figure~\ref{fig:x3x1conv} shows the axial ratio of the low-density
cavity in the Galactic plane ($x_3/x_1$) versus time for different
resolutions ($100^3, 200^3$ and $300^3$) to verify convergence (see
Section~\ref{simulations}).  From this, we confirm that convergence is
established at $200^3$ which we use for the analysis.
Figure~\ref{x1x3ratio} shows the relationship between the $x_{3}/x_{1}$
axial ratio and $\beta$ at an age of 5 Myr and 12.5 Myr. The
similarity between the four panels shows that the axial ratio of a
superbubble in the Galactic plane does not depend significantly on the
vertical stratification of the density and the vertical stratification
of the magnetic field (see also
Figures~\ref{ExpConst_B_40},~\ref{ExpEquip_B_40},~\ref{B40fig}
and~\ref{DLEquip_B_40}). The important parameters are the age of the
bubble and the strength of the magnetic field in the equatorial
plane. Even at the relatively young age of 5 Myr, a bubble in a medium
with $\beta=1$ will be elongated along the magnetic field with axial
ratio $\sim$0.8. After 12.5 Myr, the size of the bubble perpendicular
to the magnetic field is only 60\% of its size along the magnetic
field. Variation of $\beta$ by a factor of three either way changes
the predicted shape of the bubble in the Galactic plane from an axial
ratio $\sim 0.4$ to $\sim 0.8$ at 12.5 Myr.

\begin{figure}
\centerline{\resizebox{0.7\columnwidth}{!}{\includegraphics[angle=0]{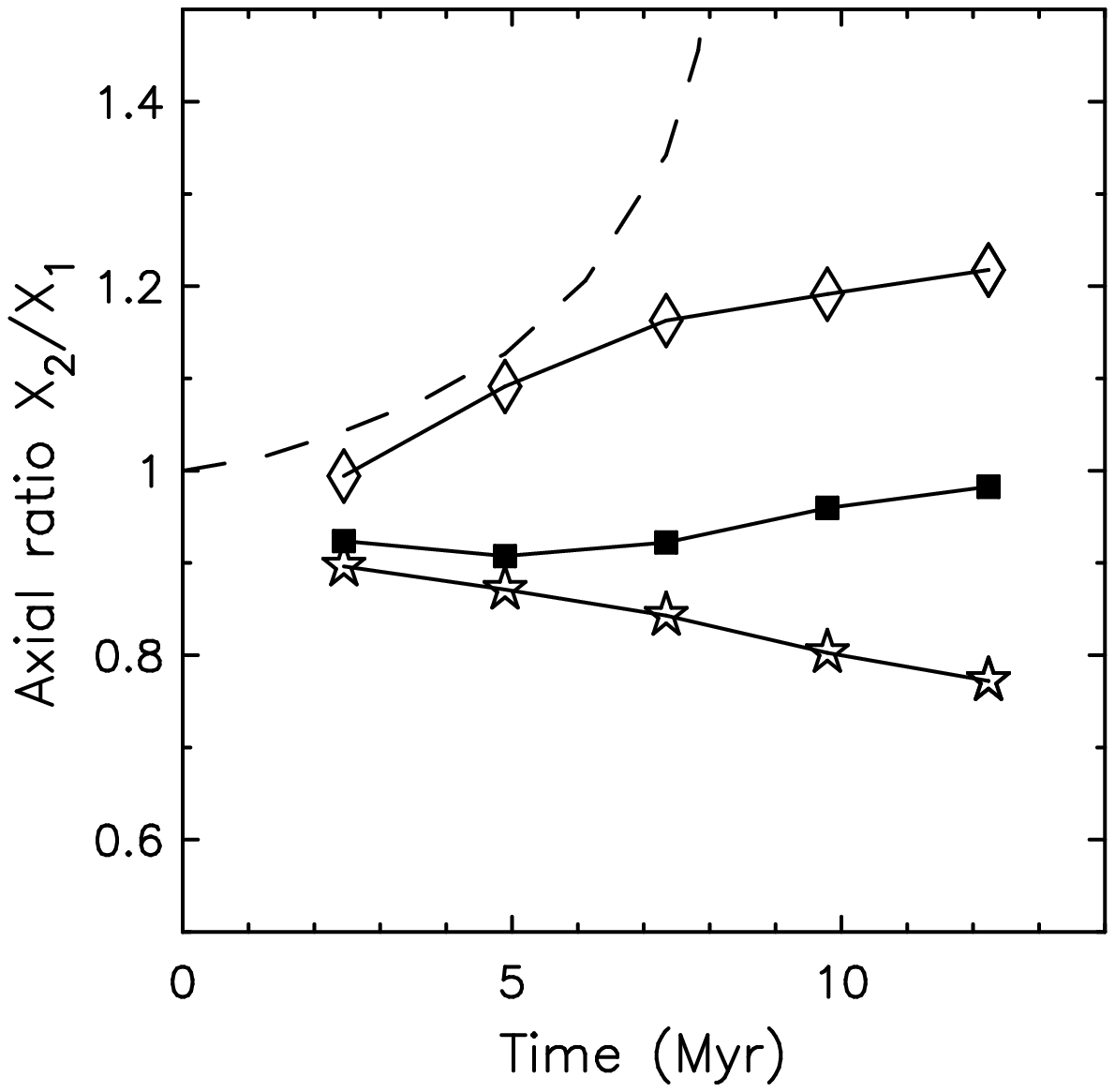}}}
\caption{Axial ratio $x_{2}/x_{1}$ as a function of time for a bubble
in an exponential atmosphere with $\beta=\infty$ (diamonds; ExpH),
$\beta=1$ with $B\sim\rho^{1/2}$ (filled squares; ExpEBc), and
$\beta=1$ with constant B (stars; ExpCBc). Also shown is the axial
ratio of the K60 model as a function of time
(dashed curve).
\label{x2x1ratioexp}
} 
\end{figure}

Figure~\ref{x2x1ratioexp} shows the time evolution of the
$x_{2}/x_{1}$ axial ratio of the low density cavity for ExpH
(diamonds), ExpCBc (stars), ExpEBc (filled squares), and the
Kompaneets model (dashed line). At early times, $t\lesssim 5$
Myr, the evolution of ExpH follows the Kompaneets model. The
difference between the ExpH simulation and the Kompaneets model at
later times is the result of inertia of the swept-up interstellar
medium.  ExpCBc and ExpEBc evolve in a $\beta=1$ medium which resists
expansion along the $x_{2}$ axis. Contrary to ExpH, at early times,
the axial ratios of ExpCBc and ExpEBc decrease, indicating that the
expansion along the $x_{1}$ axis is faster than along the $x_{2}$
axis. This trend continues for the ExpCBc simulation while it reverses
for the ExpEBc simulation where the top of the bubble accelerates at
later times.  In the ExpEB simulations, confinement in the $x_{2}$
direction by the magnetic field decreases rapidly as the bubble grows
beyond approximately one scale height for all values of $\beta$.  This
allows the top of the bubble to accelaterate at later times,
eventually leading to blow-out if the source is sufficiently strong
\citep{1998MNRAS.298..797T}.  Breakout in our ExpEBc simulation
occurs  at $t \gtrsim 17.5\ \rm Myr$.  The constant magnetic
field in all the ExpCB simulations confines the bubble in the $x_{2}$
direction, preventing blow-out.

Taking into account differences in density and luminosity of the
source, our axial ratios agree fairly well with those of
\citet{1986PASJ...38..697T}, while the results of \citet{MacLow1989}
agree better with the Kompaneets solution. Differences between our results
and those of \citet{MacLow1989} appear to be related to differences in
source luminosity and the density distribution in the atmosphere.

\begin{figure}
\centerline{\resizebox{0.7\columnwidth}{!}{\includegraphics[angle=0]{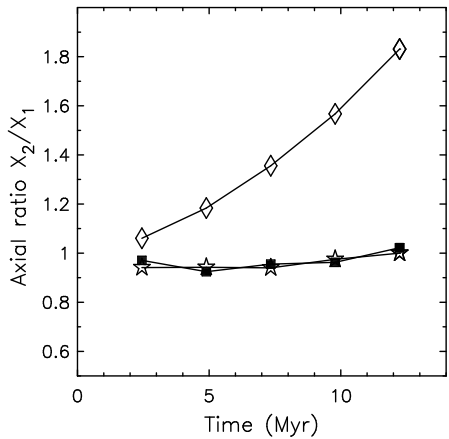}}}
\caption{Axial ratios $x_{2}/x_{1}$ as a function of time for a bubble
in the Dickey and Lockman atmosphere with $\beta=\infty$ (diamonds;
run DLH), $\beta=1$ with $B\sim\rho^{1/2}$ (filled squared;
run DLEBc), and $\beta=1$ with constant B (stars; run
DLCBc).
\label{x2x1ratioDL}
} 
\end{figure}

Figure~\ref{x2x1ratioDL} shows the time evolution of the $x_{2}/x_{1}$
axial ratio of the low density cavity for DLH (diamonds), DLCBc
(stars), and DLEBc (filled squares). The evolution of the
$x_{2}/x_{1}$ ratio of DLH shows the bubble accelerating in the
$x_{2}$ direction. The axial ratio at any time is larger than those
measured for ExpH because the bubble is expanding in both the positive
and negative $x_{2}$ directions. DLCBc and DLEBc evolve in a $\beta=1$
magnetic field. The shape of the cavity in the $x_{1}x_{2}$ plane
remains nearly circular for both magnetic field configurations, even
at later times. A stronger magnetic field ($\beta < 1$) results in a
cavity elongated along the $x_1$-axis. As for the exponential
atmosphere, the bubble in the DLEB simulations eventually breaks out,
whereas the bubbles in the DLCB simulations remain confined by the
magnetic field.

\subsubsection{Deriving Scale Height and Age From Analytic Models}
\label{comparingkomp}

The elongation of a bubble along the magnetic field has important
consequences for age estimates of the bubble from observations, which
usually assume axial symmetry or even spherical symmetry. The axial
ratio $x_{3}/x_{1}$ can be observed for superbubbles in face-on
galaxies, but not for Galactic superbubbles which are observed from
within the Galactic plane ($x_{1}x_{3}$ plane). Within the Galactic
plane, the available observations are the radius of the bubble in the
plane of the sky and the expansion velocity perpendicular along the
line of sight.  The age of the bubble derived from observations is
$t_{obs}=\alpha R_{obs}/v_{exp}$, assuming an expansion law of the
form $R\sim t^{\alpha}$. The elongated shape of the bubble along the
direction of the magnetic field implies that, on average, the
expansion velocity is larger along the $x_{1}$-axis than along the
$x_{3}$ axis. \textit{An observer who looks at the bubble along the
$x_{1}$ axis (looking along the magnetic field lines) will see the
small dimension of the bubble along the $x_{3}$ axis but measure the
larger expansion velocity along the $x_{1}$ axis. An observer who
looks at the bubble along the $x_{3}$ axis (looking perpendicular to
the magnetic field lines) will see the large dimension of the bubble
along the $x_{1}$ axis but measure the smaller expansion velocity
along the $x_{3}$ axis.} The two observers will derive significantly
different ages for the bubble from their observations. If the radius
of the bubble along the $x_{1}$ axis is $R_{0}$ and the axial ratio
$x_{3}/x_{1}$ is $q$, the age derived by observer A (looking along the
field lines) is
\begin{equation}
t_{A}=\alpha\frac{qR_{0}}{v_{exp,A}},
\end{equation}
and the age derived by observer B (looking perpendicular to the field lines) is
\begin{equation}
t_{B}=\alpha\frac{R_{0}}{v_{exp,B}}.
\end{equation}

On average, $v_{exp,B}=q v_{exp,A}$, therefore the ratio of the ages
derived by the two observers is
\begin{equation}
\frac{t_{A}}{t_{B}}=q^2.
\end{equation}
For a moderate $x_{3}/x_{1}$ axial ratio of 0.8
(Figure~\ref{x1x3ratio}), the ratio $t_{A}/t_{B}=0.64$. Age estimates
for observer A looking along the $x_{1}$ axis, and observer B looking
along the $x_{3}$ axis were derived from our simulations. The ratio of
these age estimates is even somewhat smaller than $q^{2}$, especially
for lower values of $\beta$, because the expansion of the shell
actually stalls in the $x_3$ direction after a finite amount of time.
This illustrates the importance of MHD effects on the derivation of
the age of a superbubble.

\begin{center}
\begin{table}[!ht]
\caption{Fitting the Kompaneets Model to Simulations\\ in the $x_{3}x_{2}$ plane at $t=7.3$ Myr}
\centering
\begin{tabular}{cccccccc}
\hline\hline
& \multicolumn{3}{c}{H} && \multicolumn{3}{c}{Age}\\
& \multicolumn{3}{c}{(pc)} && \multicolumn{3}{c}{(Myr)}\\
\cline{2-4}\cline{6-8}
$\tilde{y}$ & ExpH   & ExpCBc & ExpEBc   && ExpH    & ExpCBc  & ExpEBc  \\
\hline
1.4         & 106    &  78    &\nodata  && 5.39    & 3.23    &\nodata \\
1.5         &  95    &  70    &\nodata  && 5.06    & 3.04    &\nodata \\
1.6         &  85    &  62    & 69      && 4.71    & 2.79    &3.33    \\
1.7         &  75    &  55    & 61      && 4.26    & 2.54    &3.02    \\
1.8         &\nodata &\nodata & 53			&& \nodata & \nodata &2.65    \\
\hline
\end{tabular}
\label{overlaycompare}
\end{table}
\end{center}

The three-dimensional structure of a magnetized bubble not only
affects age estimates, but also estimates of the scale height of the
surrounding medium from the observed shape of the
bubble. \citet{1999ApJ...516..843B} outline a method for fitting the
Kompaneets model to observations to obtain the scale height of the
ambient medium and the age of a bubble. Our simulations can be used to
assess systematic errors introduced by applying an axially symmetric
hydrodynamic model to a magnetized superbubble. Inspection of
Figures~\ref{ExpConst_B_40}, \ref{ExpEquip_B_40}, \ref{B40fig}, and
\ref{DLEquip_B_40} shows that in our simulations the Kompaneets model
can only fit the shape of the cavity in an exponential atmosphere
looking approximately along the magnetic field lines, or the circular
shape of the cavity in a DL atmosphere looking approximately
perpendicular to the field lines. In the latter case, fitting the
Kompaneets model to the circular shape of the cavity would imply a
very large scale height. Here we focus on fitting the Kompaneets model
to the ExpH, ExpCBc, and ExpEBc simulations, in the $x_{3}x_{2}$
plane.
 
To determine a range of Kompaneets solutions that fit these bubbles,
an initial estimate for $\tilde{y}$ (Equation~\ref{ybasu}) was derived
from the $x_{3}/x_{2}$ axial ratio of the simulation. Starting with
this estimate, overlays of the Kompaneets solution were visually
fitted to the simulations to obtain a range of acceptable
$\tilde{y}$ values, with the additional constraint that the source of
the Kompaneets model must coincide with the source in the simulations to
within 2 pixels, i.e. the radius of the source in the simulations. 

Fits of the Kompaneets model have also be used to estimate the age of a
bubble, if the mechanical luminosity of the source and ambient density
are known. \citet{1999ApJ...516..843B} showed that at the level of the
source, the radius of the Kompaneets bubble is nearly equal to the
radius of a spherical bubble described by \citet{1975ApJ...200L.107C}.
From Equation~(\ref{rbasu}), at $z=0$, we find
\begin{equation}
R(z=0,\tilde{y}) = 2H \arccos\left(1-\frac{\tilde{y}^{2}}{8}\right).
\label{rbasuz0}
\end{equation}

As $\tilde{y}$ and $H$ are determined from the fit,
Equation~(\ref{rbasuz0}) can be substituted into
Equation~(\ref{rcastor}) with $L_{s}=3\times 10^{37}$ erg s$^{-1}$ and
$\rho=1.67\times 10^{-24}$ g cm$^{-3}$ to find the age of the bubble.

Table~\ref{overlaycompare} shows results of visual fits of the
Kompaneets model to the ExpH, ExpCBc, and ExpEBc simulations at an age
of 7.3 Myr, looking along the x$_{1}$ axis. Scale height and age are
listed as a function of $\tilde{y}$ for those models that provided an
acceptable fit. For all three simulations, the fitted scale height is
smaller for larger values of $\tilde{y}$, simply indicating a range of
acceptable axial ratios that fit the shape of the bubble. Simulations
ExpH and ExpCBc are fitted by the same range of $\tilde{y}$ while
ExpEBc is better fitted by higher values of $\tilde{y}$. This reflects
the more elongated shape of the cavity in ExpEBc seen in
Figure~\ref{ExpEquip_B_40}. For ExpH, we find that the scale height is
close to 100 pc. Although ExpH and ExpCBc are fitted by the same range
of $\tilde{y}$, the scale height derived from fitting ExpCBc is
smaller because the bubble is more confined. This confinement of the
bubble by the magnetic field while maintaining an elongated shape of
the cavity is the root cause of the 30\% to 50\% lower scale heights
resulting from fits of the Kompaneets model to a magnetized
bubble. Ages derived from fits to the magnetic simulations are
consistently a factor $\sim 2$ smaller than the actual age of the
simulation. We note in passing that \citet{KooMcKee1990} found that
the Kompaneets model blows out a factor $\sim 2$ earlier than their
numerical hydrodynamic solutions.

As discussed earlier, the bottom of the bubble in the
K60 model does not penetrate deeper than 1.4
exponential scale heights below the level of the source. If a bubble
would penetrate deeper into the atmosphere, an observer could conclude
that the Kompaneets model is not applicable.  However, none of our
simulations penetrate as deep as the limiting value of 1.4 scale
heights below the level of the source. In practise, this would not
consitute a conclusive test whether or not the Kompaneets model can be
applied.

\section{APPLICATION TO THE W4 SUPERBUBBLE}
\label{app}

As a specific application of our simulations we investigated the
result of \citet{1999ApJ...516..843B} that the scale height of the
interstellar medium around the W4 superbubble
\citep{1996Natur.380..687N} is as small as 25 pc, approximately a
factor 4 smaller than the canonical value of 100 pc.
\citet{komljenovic1999} investigated this issue with two-dimensional
MHD simulations, and suggested that the elongated shape of the bubble
could only be generated by a magnetic field perpendicular to the
Galactic plane (see also West et al. 2007). Our simulations allow us
to consider the elongated shape of the bubble seen by an observer
looking along the magnetic field lines.  This perspective is not
possible in two-dimensional simulations.  This section is organized as
follows. First we review some basic properties of the W4 region,
because we found that the ambient density is a factor $\sim 2.5$ lower
than adopted by previous authors.  Second, we proceed with our
analysis of the scale height near W4.

\citet{1996Natur.380..687N} showed that an expanding shell in the
Perseus arm associated with the Cas OB6 association
\citep{braunsfurth1983} is a conical cavity with no apparent upper
boundary in \HI. This structure was interpreted as a ``chimney'', a
large stellar wind bubble which has broken out of the Galactic disk
and expands into the Galactic halo. Evidence for a vertical flow
inside the cavity comes from elongated \HI\ structures associated with
a molecular cloud \citep{1996Natur.380..687N}, and a comet-shaped
molecular cloud \citep{1996Natur.380..687N,heyer1996}. The lower part
of the bubble wall is bright in H$\alpha$ and is also seen in radio
continuum images \citep{1996Natur.380..687N}.  The bright \HII\ region
at the bottom of the cavity is known as W4 \citep{westerhout1958}. The
bubble appears to be confined at the bottom by a dense cloud.

The probable source of the bubble is the star cluster OCL~352
(IC~1805), which is located near the bottom of the
chimney. \citet{1996Natur.380..687N} list 9 O stars in the cluster, with
the earliest spectral type O4I for the star BD+60504. The presence of
the O4 star in OCL~352 makes it likely that no supernova explosion has
yet occured in the cluster, so that the bubble is a true stellar wind
bubble.  Most recent age determinations for the cluster are in the
range 1 to 3 Myr \citep{hillwig2006,rauw2004,1995ApJ...454..151M}, but ages up
to 5 Myr have been suggested \citep{kharchenko2005}.  The kinetic
energy released in the stellar wind of the stars in the cluster is $
L_{mech} = 3 \times 10^{37}\ \rm erg s^{-1}$ \citep{1996Natur.380..687N},
most of which is emitted by BD$+$60504 and two O5 stars.

\citet{dennison1997} reported an elongated H$\alpha$ shell surrounding
the \HI\ cavity that appears closed $6\degr$ (230 pc) above the star
cluster, outside the field of view of the original \HI\ images of
\citet{1996Natur.380..687N}. Extended \HI\ images that cover the top
of the H$\alpha$ shell do not show a cap in the neutral gas
\citep{normandeau2000}.  The H$\alpha$ shell was interpreted as the
inner wall of the superbubble ionized by the stars in OCL~352.
\citet{dennison1997} found that the age of the superbubble exceeds the
maximum possible age of OCL~352 by a factor $\sim 3$, using the
mechanical luminosity of the star cluster ($L_{mech} = 3 \times
10^{37}\ \rm erg s^{-1}$) and the density of the ambient medium ($n_0
= 5\ \rm cm^{-3}$) given by \citet{1996Natur.380..687N}.  The large
dynamical age of the bubble was taken as an indication that the star
cluster OCL~352 by itself may not be the only source of the
superbubble. \citet{1999ApJ...516..843B} found that the age of the
bubble is consistent with the age of the star cluster by applying the
K60 model to the W4 superbubble, assuming a scale height of only 25 pc
for the ambient medium.  However, \citet{dennison1997} noted that a
lower ambient density could decrease the dynamical age derived for the
W4 superbubble.

The original \HI\ number density was derived by
\citet{1996Natur.380..687N} from an \HI\ column density map.  This map
was obtained by integration of the \HI\ brightness temperature over a
velocity range of $14.8\ \kms$. The ambient density was then
calculated by dividing the \HI\ column density adjacent to the cavity
by the diameter of the bubble, assuming that all the \HI\ emission is
local to the W4 superbubble.  This assumption leads to a significant
over-estimate of the ambient density, because \HI\ emission in the
velocity range of the bubble originates from a much longer section of
the line of sight.  We re-analyzed the \HI\ data of the W4 region, now
publicly available as part of the Canadian Galactic Plane Survey
\citep{taylor2003}, to re-determine the ambient density.

The \HI\ column density inside the cavity of the W4 superbubble,
avoiding the \HI\ features known to be inside the cavity
\citep{1996Natur.380..687N}, is $1.1 \times 10^{21}\ \rm cm^{-2}$.
The medium inside the superbubble is expected to be a highly ionized
tenuous plasma at a temperature of a few million Kelvin. The ionized
bubble wall seen on both sides of the cavity in radio continuum and
H$\alpha$ emission shows that the interior of the bubble is optically
thin to Lyman continuum photons.  \citet{dennison1997} and
\citet{1999ApJ...516..843B} assume this to be true in their analysis
of the ionization of the bubble wall by the star cluster. Therefore,
\HI\ observed inside the cavity must be located in the foreground and
the background. Some of this \HI\ emission could in principle belong
to the bubble wall if the expansion of the superbubble had
stalled. There is no reason to believe that the expansion of the
bubble has stalled because the source is still active. 

Although the evidence that the \HI\ steamers identified by
\citet{1996Natur.380..687N} are located inside the cavity of the W4
bubble is compelling, the \HI\ emission that seems to partly fill the
cavity everywhere is likely unrelated gas, that was included in the
density estimate of \citet{1996Natur.380..687N}. If the \HI\ column
density inside the cavity is subtracted, the density of the ambient
medium is reduced to $\sim 2\ \rm cm^{-3}$ assuming the same
line-of-sight dimension of the bubble. A better estimate of the
density of the ambient medium is obtained by dividing the mass of the
shell of swept-up gas by the volume of the cavity. This is impractical
for the W4 bubble because there is no clear limb-brightened \HI\ shell
visible in the column density map. \citet{1999ApJ...516..843B} found a
higher density of 10 $\rm cm^{-3}$ from their analysis of the shape of
the ionization front, which depends on the details of their model.
Although the lower value of the ambient density derived from the \HI\
data may not resolve the age problem altogether, we follow the
suggestion by \citet{dennison1997} that the location of the cluster is
strong evidence for its identification as the source of the
superbubble.

We now return to the issue of the small scale height found by
\citet{1999ApJ...516..843B}. \citet{normandeau2000} found no
evidence for such a small scale height in \HI\ images of the
region, but, as noted before, the \HI\ images are contaminated
to some extent by emission in the foreground and background. The
presence of molecular gas with a small scale height may have
contributed to the result obtained by \citet{1999ApJ...516..843B}, but
the molecular gas traced by the CO line appears too fragmented to
confine the shape of the bubble.  The scale height found by fitting
the Kompaneets model to the observed shape of the bubble may also be
artificially small by a factor 2 (see \S \ref{comparingkomp})
for an observer looking along the magnetic field parallel to the
Galactic plane. This factor is not enough to explain the very small
scale height derived by \citet{1999ApJ...516..843B}. Also, the age
derived by \citet{1999ApJ...516..843B} would be too small by a factor
2, further increasing the discrepancy between the dynamical age of the
bubble and the age of the star cluster.  None of our simulations
evolve into a cavity that is as narrow as the W4 super bubble.

\section{FARADAY ROTATION BY A MAGNETIZED BUBBLE}

Polarimetric observations of the Galactic plane at arcminute
resolution \citep{taylor2003,mcclure2005} allow measurements of
Faraday rotation of diffuse Galactic synchrotron radiation and
background compact extragalactic sources to probe the magneto-ionic
interstellar medium down to parsec scales.  The plane of polarization
of linearly polarized synchrotron emission rotates by an angle $\Delta
\theta$ (radians) as the radiation passes through a magnetized plasma
according to $\Delta \theta = \lambda^2 RM$, with $\lambda$ the
wavelength of the radiation in meters. The line-of-sight integral
\begin{equation}
RM = 0.812 \int n_{e} {\bf B \cdot dl}  \label{RMequation}
\end{equation} 
is the rotation measure (radians $\rm m^{-2}$), with $n_e$ the
electron density ($\rm cm^{-3}$), and ${\bf B \cdot dl}$ the magnetic
field vector $B$ ($\mu\rm G$) projected onto the line of sight (in
parsec).  The rotation measure can be determined observationally by
observing $\Delta \theta$ at different, but closely spaced wavelengths.
Structure along the line of sight may create a measurable rotation of
the plane of polarization, but it may not have a sufficiently large
emission measure for any thermal emission to be detected.

\begin{figure*}
\resizebox{\textwidth}{!}{\includegraphics[angle=-90]{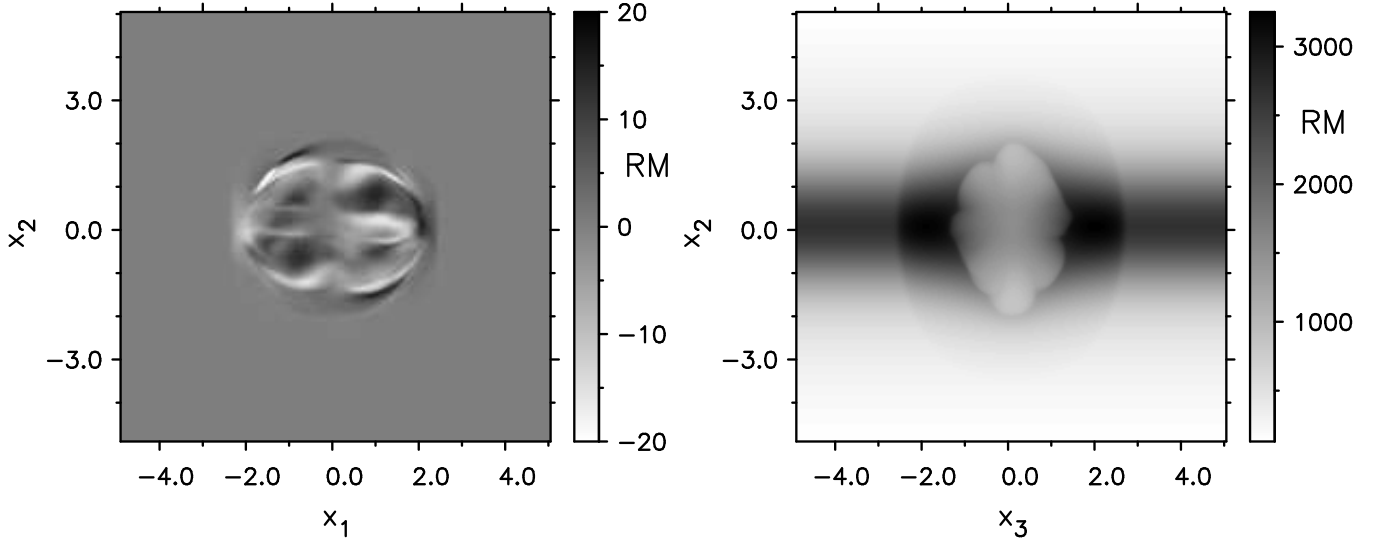}}
\caption{ Images of rotation measure of the simulation DLCBc at age 10
Myr (see also Figure~\ref{B40fig}) along a direction perpendicular to
the undisturbed Galactic magnetic field (left) and along the direction
of the magnetic field (right). Grayscales are linear from $-20$ to
$+20$ radians $\rm m^{-2}$ (left) and from 130 to 3252 radians $\rm
m^{-2}$ (right).  Wrapping of the magnetic field lines around the
low-density cavity creates significantly different Faraday rotation
signatures for the same bubble seen from different directions.  If the
line of sight is perpendicular to the Galactic magnetic field, the
shell of swept-up interstellar medium is almost invisible in the
rotation measure map. If the line of sight is along the Galactic
magnetic field, the largest rotation measures are associated with the
(thick) shell. The unit of length on the axes is 100 pc.
\label{RM-fig}
}  
\end{figure*}

Equation~(\ref{RMequation}) cannot be inverted to obtain a unique
solution to the magnetic field structure, even if independent
information on the electron density $n_e$ along the line of sight is
available.  Instead, the data can be interpreted by line-of-sight
integration of a model distribution of the density and magnetic field,
which depends critically on the assumed distributions of the electron
density and magnetic field.  If an ad hoc model for density and
magnetic field is adopted to interpret the observations, careful
consideration should be given to basic questions regarding
self-consistency, uniqueness and dynamical properties of the proposed
model. Our three-dimensional MHD simulations of magnetized
superbubbles constitute a significant step forward in the modeling of
Faraday rotation by Galactic superbubbles. In this paper, we only
consider the bubble as a Faraday screen that does not emit synchrotron
emission. This situation applies in particular to compact polarized
sources in the background.

The simulations provide a self-consistent set of models that can be
used to calculate the rotation measure imposed on background emission
by a magnetized superbubble. The non-spherical evolution of the
superbubble described in Section~\ref{simulations} may have very
significant effects on the predicted rotation measure. The effect of
nearby superbubbles on Faraday rotation of background sources was
first discussed by \citet{vallee1993}.  In this paper we discuss only
the Faraday rotation signature of distant bubbles, for which we can
assume that all lines of sight are parallel.  We assume here that the
bubble is completely ionized. This allows a first exploration of the
general effects of the topology of the density and magnetic field,
without introducing more parameters.

Figure~\ref{RM-fig} shows the rotation measure for a line of sight
along the $x_{3}$ axis (looking perpendicular to the Galactic magnetic
field) and the $x_{1}$ axis (looking along the Galactic magnetic
field) of the simulation DLCBc (Table~\ref{simsrun}) at an age of 10
Myr (Figure~\ref{B40fig}). The line-of-sight integration of
Equation~(\ref{RMequation}) was done through the entire simulated
volume as shown in Figure~\ref{B40fig}. It includes some undisturbed
medium in the foreground and the background of the bubble.

Comparing the rotation measure maps with the density and magnetic
field structure (Figure~\ref{B40fig}) reveals some general
characteristics of the Faraday rotation of all bubbles in our
simulations. As the magnetic field is pushed aside, the field lines
wrap around the expanding cavity. The largest amplification and the
largest change of direction of the magnetic field occur just outside
the cavity, in the equatorial plane (for the DL density
distribution). The density is also enhanced, compared with the
surrounding undisturbed interstellar medium, but by different amounts
depending on location in the shell. Simulations with a small scale
height of the Galactic magnetic field ($B \sim \rho^{1/2}$) show this
compression also at the top of the bubble, whereas simulations with a
large scale height of the magnetic field (constant B) do not. The
largest Faraday rotation therefore occurs in a relatively thin region
around the cavity, being strongest close to the Galactic plane (see
Figure~\ref{RM-fig}). The Faraday rotation inside the cavity is much
smaller by comparison, because of the low density and the low chaotic
magnetic field there.

If the line of sight is perpendicular to the direction of the
undisturbed magnetic field, the near and the far side of the bubble
wall have rotation measures up to $\sim$100 radians $\rm m^{-2}$, with
opposite signs. These mostly cancel each other, but asymmetries
between the front side and the back side of the shell create structure
in the rotation measure distribution. Asymmetries arise from the
instabilities in the shell mentioned in
Sections~\ref{ExpCBandE}~\&~\ref{analysis}. The strongest signal is
expected from asymmetries that occur in locations where a perfectly
symmetric bubble would have bent the Galactic magnetic field
sufficiently to create a significant line of sight component of the
magnetic field. As the magnetic field is tightly wrapped around the
cavity, the highest values of the rotation measure resulting from
front-to-back asymmetries are therefore expected for lines of sight
that intersect the cavity, away from the vertical axis that intersects
the source.  Our three-dimensional simulations contain such
asymmetries, resulting in the structures seen in Figure~\ref{RM-fig}
with amplitudes of the order of 20 radians $\rm m^{-2}$. 

\begin{figure*}
\resizebox{\textwidth}{!}{\includegraphics[angle=-90]{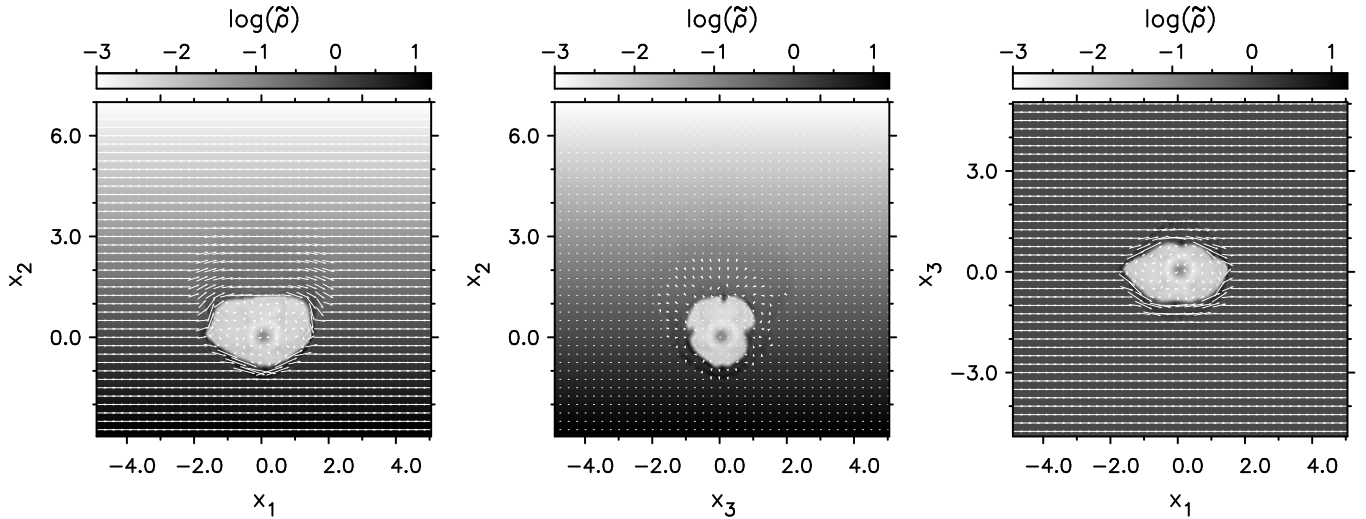}}
\caption{Simulation ExpCB with cooling at an age of 10 Myr for
comparison with Figure~\ref{ExpConst_B_40}. Magnetic field initially
oriented along x$_{1}$ axis, $\beta = 1$. Panels show slices through
the cube at the location of the source,
($x_{1}$,$x_{2}$,$x_{3}$)=(0,0,0), in three orthogonal
planes. Grayscales show the gas density on a logarithmic scale from
$10^{-3}$ (white) to $10^{1.2}$ (black) cm$^{-3}$. The vector field
depicts the projection of magnetic field vectors on the plane of the
image. The unit of length on the axes is 100 pc.
\label{ExpConst_B_40_cool}
} 
\end{figure*}

\begin{figure*}
\resizebox{\textwidth}{!}{\includegraphics[angle=-90]{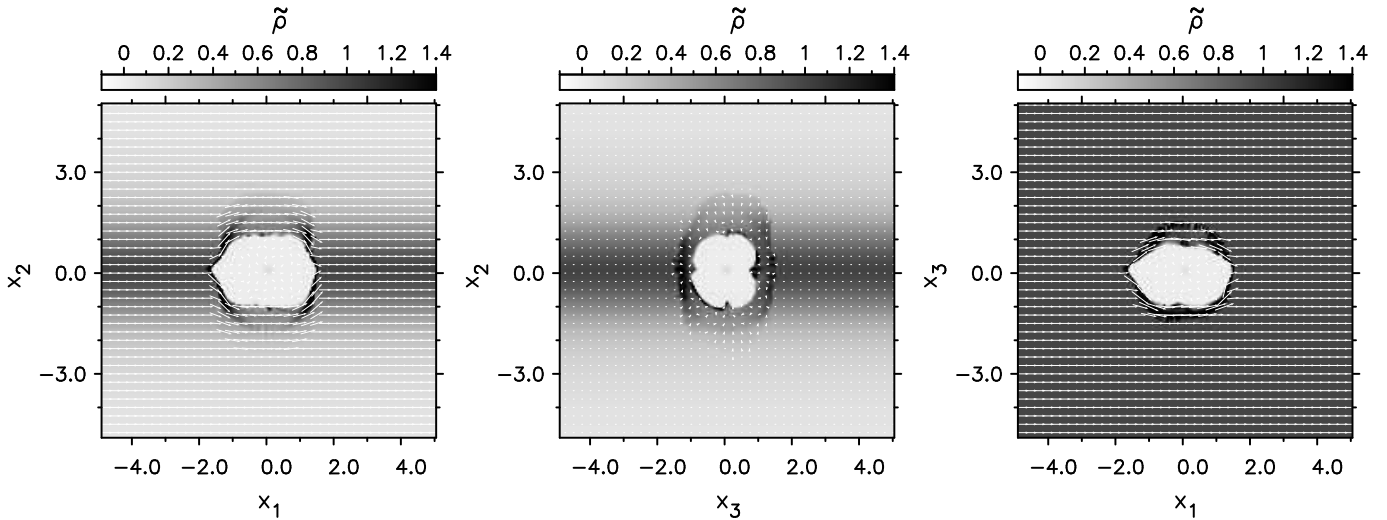}}
\caption{Simulation DLCB with cooling at an age of 10 Myr for
comparison with Figure~\ref{B40fig}. Magnetic field initially
oriented along x$_{1}$ axis, $\beta = 1$. Panels show slices through
the cube at the location of the source,
($x_{1}$,$x_{2}$,$x_{3}$)=(0,0,0), in three orthogonal
planes. Grayscales show the gas density on a logarithmic scale from
$10^{-3}$ (white) to $10^{1.2}$ (black) cm$^{-3}$. The vector field
depicts the projection of magnetic field vectors on the plane of the
image. The unit of length on the axes is 100 pc.
\label{DLConst_B_40_cool}
} 
\end{figure*}

If the line of sight is parallel to the direction of the undisturbed
magnetic field, rotation measures are much higher everywhere than in
the previous case.  The front and back side of the shell reinforce
each other, and high rotation measures are expected for lines of sight
that intersect the shell. The top of the bubble increases the rotation
measure a few hundred parsecs above the Galactic plane by $\sim 20\%$
compared with the undisturbed atmosphere in Figure~\ref{RM-fig}. For
the simulation DLEBc (small magnetic scale height), the increase is
$\sim 30\%$ compared with the undisturbed atmosphere. Within 100 pc
from the Galactic plane, the rotation measure varies with location
because of the bubble, but the mean of the rotation measure across the
bubble is almost the same as the mean rotation measure of the
undisturbed atmosphere. Here, the effect of the bubble is to create a
significant variation in rotation measure for different lines of
sight, with little effect on the mean rotation measure taken over a
large area. Super bubble shells may therefore enhance the rotation measure 
in the disk-halo interface and affect estimates of the scale height of the 
Galactic magnetic field from rotation measure analysis.

\citet{vallee1983} reported a large magnetized bubble associated with
the Gum nebula, based on a number of high rotation measures towards
extragalactic sources. Recently, \citet{stil2007} found this bubble in
an all-sky image of depolarization of extragalactic sources, that was
associated with excess rotation measure. This bubble stands out in the
data because of the high rotation measure in the shell. The line of
sight in this case is nearly parallel to the magnetic field in the
local spiral arm.

\section{THE EFFECTS OF COOLING}
\label{cooling-sec}

So far we have discussed adiabatic simulations of super bubbles as an
approximation that energy losses are compensated by an internal or an
external radiation field. In this section we briefly discuss
simulations with cooling in the two main density stratifications
discussed in this paper: the exponential atmosphere and the Dickey \&
Lockman atmosphere.  The cooling represents an extra term in the
energy equation (10) and is implemented as an explicit source term in
the code. As with all physical processes, there is a maximum allowed
time step for numerical stability associated with the cooling; in our
case it is reset to 10\% of the radiative cooling time when it is
below the normal hydrodynamic CFL condition.

The cooling function by \citet{raymond1976} and \citet{dalgarno1972},
using the abundances of \citet{allen1973}, was implemented into
zeus-mp in the form of analytic expressions adopted by
\citet{tomisaka1981}. Although some abundances were revised downwards
later, the difference between the cooling rates adopted in our
simulations and the solar abundance cooling curve of \citet{gnat2007}
is smaller than other sources of uncertainty introduced by finite
resolution, and ionization by the central star cluster. If an accuracy
better than a factor $\sim 3$ is required in the cooling rate, the
location of the superbubble in the Galaxy also becomes important
because of variation in the abundances with location in the Galaxy
\citep[e.g.][]{rudolph2006}.

The radiatively cooled simulations were done in atmospheres maintained
at a temperature of 8000 K, with the same resolution as the adiabatic
simulations discussed in Section~\ref{simulations}, and with the same
mechanical luminosity of the source. In these simulations, cooling
begins at a time $t = 4$ Myr, soon after the first supernova
explosion, at the beginning of the radiative phase following
\citet{mccray_kafatos1987}. The effects of cooling on the shape of the
cavity are evaluated at $t = 10$ Myr.

\subsection{Results with cooling}

Figure~\ref{ExpConst_B_40_cool} and Figure~\ref{DLConst_B_40_cool}
show simulations of a bubble with cooling in an exponential atmosphere
and in a Dickey \& Lockman atmosphere at an age og 10 Myr. These
figures can be compared with the adiabatic simulations shown in
Figure~\ref{ExpConst_B_40} and Figure~\ref{B40fig}.  Qualitatively the
features in these simulations are very similar.  The cavity is
elongated along the direction of the magnetic field.  The shell is
thicker in directions perpendicular to the magnetic field in the
undisturbed atmosphere ($x_2$ and $x_3$), while significant
compression of the swept-up material in the direction of the magnetic
field leads to a thinner, denser shell. Closer inspection shows that
the cavities of both cooled simulations are more elongated in the
direction of the magnetic field ($x_1$). The shell also seems to
display stronger instabilities, leading to irregularities in the shape
of the cavity, and increased scatter in the derived axial ratios of
the cavity. As expected, the simulations with cooling lead to shells
that are thinner, with a higher density and stronger magnetic field,
than the adiabatic simulations. The effect of resolution on the cooled
regions is to smear out small-scale high-density structure with the
shortest cooling time and the lowest
temperature. \citet{deavillez2004} found convergence in the maximum
density and lowest temperature in their simulations for a grid size
$\lesssim 1.1$ pc. The density and temperature inside the swept-up
shell in our simulations are probably also affected by resolution, but
this paper focuses on the shape of the bubble, not the conditions
inside the shell. \citet{garcia2000} found that the shape of planetary
nebulae in their simulations was not affected by resolution.

\begin{figure}
\resizebox{\columnwidth}{!}{\includegraphics[angle=0]{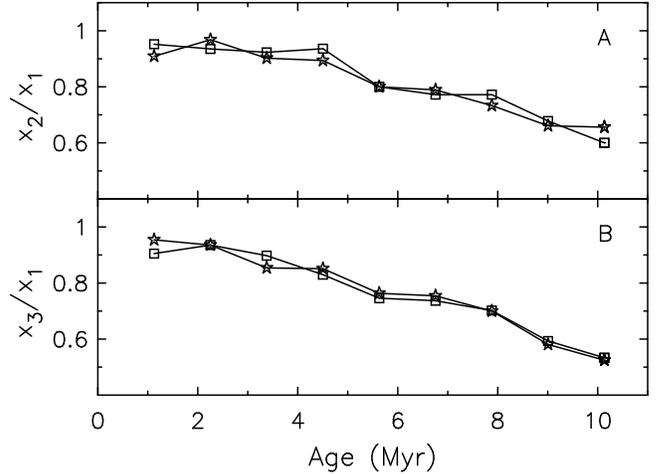}}
\caption{ Axial ratio $x_2/x_1$ (A) and $x_3/x_1$ (B) as a function of
the age of the bubble for cooled simulations in an exponential
atmosphere (stars) and a Dickey \& Lockman atmosphere (squares).
Panel (A) may be compared with Figures~\ref{x2x1ratioexp} and
\ref{x2x1ratioDL}, and panel B may be compared with
Figure~\ref{fig:x3x1conv}.
\label{axial_ratio_cool-fig}
} 
\end{figure}

Figure~\ref{axial_ratio_cool-fig} shows the evolution of the axial
ratios as a function of time in these simulations.  We find that the
axial ratio $x_2/x_1$ of a bubble in the Dickey \& Lockman atmosphere
is most affected by cooling (compare
Figure~\ref{axial_ratio_cool-fig}a with Figures~\ref{x2x1ratioDL} and
\ref{x2x1ratioDL}). In the adiabatic simulation, the cavity remains
nearly circular to an age of 10 Myr, while in the cooled simulation
the cavity is more elongated in the direction of the magnetic field.
Note the change in evolution of the $x_2/x_1$ axial ratio just after
the onset of cooling.  The effect of cooling on the axial ratio
$x_2/x_1$ compared with the adiabatic case is less strong for the
exponential atmosphere, but here too the cavity becomes more elongated
in the direction of the magnetic field. A stronger elongation is also
seen in the $x_3/x_1$ axial ratio (compare
Figure~\ref{axial_ratio_cool-fig}b with Figure~\ref{fig:x3x1conv}),
but the effect is smaller than on the axial ratio $x_2/x_1$.

The observed change in axial ratios in the presence of cooling is most
likely the result of increased magnetic tension exerted by the
stronger magnetic field in the dense shell that surrounds the cavity.
This restricts the expansion perpendicular to the field ($x_2$ and
$x_3$) more in the cooled simulations than in the adiabatic
simulations. Coupling between he magnetic field and the neutral shell
is maintained through a small ionized fraction in the shell, and
ion-neutral collisions. A complete parameter study of simulations with
cooling is beyond the scope of this paper.

At least some superbubbles have a substantial fraction of the mass of
the shell ionized by the stars inside. In simulations with cooling,
the cooling time of the shell is very short compared to the age of the
bubble, and a dense neutral shell forms.  Ideally, radiative transport
of the ionizing flux of the star cluster and any unrelated massive hot
stars in the simulation volume should be included in the
simulations. Solving the radiative transport every time step in the
simulations is a computational challenge. The mean free path of an
ionizing photon is substantially smaller than the resolution of the
simulations. The ionizing flux decreases rapidly with time as the most
massive stars explode as a supernova after $\sim$ 3 Myr, and the shell
will eventually cool and become neutral.  The implicit assumption in
this paper by applying adiabatic evolution to this problem is that
heating by photo-ionization is balanced by cooling. While this is a
simplification, the observed ionization of superbubble walls suggests
that heating by ionization of the shell is a non-negligible term in
the energy equation of the shell that is not included in simuations
with cooling only.

\section{CONCLUSIONS}
\label{conc}

We present three-dimensional MHD simulations of a superbubble evolving
in a magnetized medium. In these simulations, we assume two different
atmospheric models, exponential and Dickey \& Lockman (1990), and two
different magnetic field configurations, constant magnetic field and
constant $\beta$, for varying values of the magnetic field
strength. With these simulations we aim to study the importance of MHD
effects on the interpretation of observed superbubbles.

As noted before by \citet{1998MNRAS.298..797T}, a superbubble in a
magnetized medium becomes significantly elongated along the magnetic
field. We find that the axial ratio of the bubble in the Galactic
plane is $\sim$ 0.8 after 5 Myr and $\sim 0.6$ after 12.5 Myr,
depending on the age of the bubble and the strength of the magnetic
field but not on the vertical structure of either the magnetic field
or the gas. The elongated shape of the bubble in the Galactic plane
may lead to significant errors in the age determination of magnetized
superbubbles from observations using symmetric hydrodynamic
models. The derived age depends on the location of the observer. The
ratio of the smallest age to the largest age derived by observers
assuming axial or spherical symmetry is found to be proportional to
the square of the axial ratio of the bubble in the Galactic plane,
which creates a discrepancy in the age of up to a factor $\sim$4.

We have analyzed systematic errors in the age of the bubble and the
scale height of the ambient medium introduced by fitting the
Kompaneets model to a magnetized superbubble looking along the
magnetic field lines. The scale height may be underestimated by 30\%
to 50\% and the age by 50\%.  In particular, we investigated the
curiously small scale height of the interstellar medium near the W4
superbubble found by \citet{1999ApJ...516..843B}. We re-analyzed HI
data from the Canadian Galactic Plane Survey and found that the
density of the ambient medium $n_{H}\approx 2 \rm{cm}^{-3}$ which is
smaller than previously thought. This lower density helps to diminish
the apparent age discrepancy between the W4 bubble and the star
cluster OCL 352. However, our analysis of the systematic errors
introduced by fitting the Kompaneets model shows that they are not big
enough to explain the scale height of 25 pc found by
\citet{1999ApJ...516..843B} for the W4 region.

We use the MHD simulations to predict the rotation measure
distribution of superbubbles based on three-dimensional MHD
simulations, and emphasize the importance of such simulations to make
these predictions. As expected, the appearance of a magnetized
superbubble depends on the perspective of the observer. If the
observer looks along the magnetic field lines, the largest rotation
measures are seen at the intersection of the shell with the Galactic
plane. The rotation measure is increased at larger distances from the
Galactic plane. If an observer in the Galactic plane looks
perpendicular to the magnetic field lines, the rotation measures are
much smaller, but most importantly most of the structure in rotation
measure appears in projection on the low-density cavity, and not on
the shell surrounding it.

The simulations and analysis presented in this paper highlight the
importance of three-dimensional MHD simulations of superbubbles
evolving in the Galactic magnetic field to the interpretation of new
high-resolution images of the Galactic plane at radio wavelengths from
the International Galactic Plane Survey.

\acknowledgements This research has been enabled by the use of WestGrid
computing resources, which are funded in part by the Canada Foundation
for Innovation, Alberta Innovation and Science, BC Advanced Education,
and the participating research institutions. WestGrid equipment is
provided by IBM, Hewlett Packard and SGI

\appendix

\section{The Kompaneets Solution at Early Times}

We show that the Kompaneets solution at early times can be
approximated as an expanding sphere with a center that moves upward
from the location of the source. This analytic derivation explains
quantitatively the behavior of the Kompaneets solution in
Figure~\ref{kompaneetsfig}.

The center of the bubble along the vertical axis follows from the expressions
for the top and bottom of the bubble (Equation~\ref{zupperlower})
\begin{equation}
z_c = {z_1 + z_2 \over 2} = -H \ln\bigl[1-({\tilde{y}/2)^2} \bigr].
\label{komp_center}
\end{equation}
The half-diameter of the Kompaneets solution in the vertical direction
is 
\begin{equation}
R_z = {z_1 - z_2 \over 2} = -H \ln \bigl[ { {1-\tilde{y}/2} \over {1+\tilde{y}/2}} \bigr]. 
\label{komp_radius_z}
\end{equation}
The maximum radius $R_h$ of the Kompaneets solution in a direction
perpendicular to the density gradient was given by
\citet{Bisnovatyi1995} and \citet{1999ApJ...516..843B},
\begin{equation}
R_h = 2 H \arcsin(\tilde{y}/2).
\end{equation}
The Taylor expansion of $z_c$, $R_z$, and $R_h$ in $x = \tilde{y}/2$
to third order in $x$ is
\begin{equation}
z_c = H x^2 + O(x^4),
\end{equation}
where $O({x^4})$ is the remainder that contains terms of order $x^4$ or higher.
Similarly, we have
\begin{equation}
R_z = 2 H x + {2 \over 3} H x^3 + O(x^5),
\end{equation}
and
\begin{equation}
R_h = 2 H x + {1 \over 3} H x^3 + O(x^5),
\end{equation}
so the difference $R_z - R_h$ is of order $x^3$
\begin{equation}
R_z - R_h = {1 \over 3} H x^3 + O(x^5).
\end{equation}
We see that for small $x$, the shift in the centre is $H x^2$, which
is of second order in $x$. The difference $R_z - R_h$ is of third
order in $x$.  The Kompaneets solution for small $\tilde{y}$ can
therefore be approximated by a spherical bubble that rises in the
atmosphere. The center according to equation~\ref{komp_center} is
shown in Figure~\ref{kompaneetsfig}. The offset is clear, where the
Kompaneets model is still visually a circle.  Although both the
numerical simulation and the Kompaneets model have an axial ratio that
is near unity, the difference between the Kompaneets model and the
numerical simulation is in the fact that the former is not centered on
the source.

%\begin{acknowledgements}
%\end{acknowledgements}

%\clearpage

{}

\end{document}